\documentclass[jorunal]{IEEEtran}
\usepackage[short]{optidef}
\usepackage{amsmath,amsfonts}
\usepackage{algorithm}
\usepackage{algpseudocode}
\usepackage{array}
\usepackage[caption=false,font=normalsize,labelfont=sf,textfont=sf]{subfig}
\usepackage{textcomp}
\usepackage{stfloats}
\usepackage{url}
\usepackage{verbatim}
\usepackage{graphicx}
\usepackage{cite}
\usepackage{cancel}
\usepackage[inline]{enumitem}
\usepackage{orcidlink}
\usepackage{wasysym}
\usepackage{amssymb}
\usepackage{xcolor}  
\usepackage{mathtools,relsize,stackengine}  
\definecolor{darkgreen}{RGB}{20, 120, 20}
\definecolor{darkorange}{RGB}{204, 82, 20}
\definecolor{gprop}{RGB}{2, 158, 115}

\DeclareMathSymbol{\shortminus}{\mathbin}{AMSa}{"39}
\usepackage{tikz, ifthen}
\usetikzlibrary{3d}          
\usetikzlibrary{perspective} 
\makeatletter
\tikzoption{canvas is plane}[]{\@setOxy#1}
\def\@setOxy O(#1,#2,#3)x(#4,#5,#6)y(#7,#8,#9)%
  {\def\tikz@plane@origin{\pgfpointxyz{#1}{#2}{#3}}%
   \def\tikz@plane@x{\pgfpointxyz{#4}{#5}{#6}}%
   \def\tikz@plane@y{\pgfpointxyz{#7}{#8}{#9}}%
   \tikz@canvas@is@plane
  }
\newdimen\@IEEEnormalsizeunitybaselineskip
\@IEEEnormalsizeunitybaselineskip\baselineskip
\def\normalsize{\@setfontsize{\normalsize}{10}{12.00pt}}
\normalsize
\belowdisplayskip \abovedisplayskip
\def\small{\@setfontsize{\small}{9}{10pt}}
\def\footnotesize{\@setfontsize{\footnotesize}{8}{9pt}}
\def\scriptsize{\@setfontsize{\scriptsize}{7}{8pt}}
\def\tiny{\@setfontsize{\tiny}{5}{6pt}}
\def\sublargesize{\@setfontsize{\sublargesize}{11}{13.4pt}}
\def\large{\@setfontsize{\large}{12}{14pt}}
\def\Large{\@setfontsize{\Large}{14}{17pt}}
\def\LARGE{\@setfontsize{\LARGE}{17}{20pt}}
\def\huge{\@setfontsize{\huge}{20}{24pt}}
\def\Huge{\@setfontsize{\Huge}{24}{28pt}}
\makeatother
\tikzset
{
  linea/.style={draw=gray},
  lineb/.style={draw=gray},
}

\newcommand{\drawgrid}[9] 
{
\foreach \i in {1,...,#4}
  \foreach \j in {1,...,#5}
  {
\fill[draw=black, fill=#9, canvas is #1 plane at #2=#3+#8] (\i+#6-1, \j+#7-1) rectangle (\i+#6, \j+#7);
}
}

\newcommand{\drawcube}[6] 
{
  \drawgrid{xy}{z}{#3}{#1}{#2}{#4}{#5}{#6}{white}
  \drawgrid{xz}{y}{0}{#1}{#3}{#4}{#6}{#5}{gray!10}
  \drawgrid{yz}{x}{0}{#2}{#3}{#5}{#6}{#4}{gray!40}
}

\newcommand{\simplegraph}[5] 
{
  \ifthenelse{#4=1}{\def\fillcolorA{gray}}{\def\fillcolorA{red}}
  \ifthenelse{#5=1}{\def\fillcolorB{gray}}{\def\fillcolorB{red}}
  \ifthenelse{#4=0}{\draw [black, ultra thick] (#1, 4.3+#2, 4.3+#3) -- (#1, 1.3+#2, 1.3+#3);}{}
  \ifthenelse{#5=0}{\draw [black, ultra thick] (#1, 4.3+#2, 4.3+#3) -- (#1, 7.3+#2, 4.3+#3);}{}
  \ifthenelse{#5=0}{\draw [black, ultra thick] (#1, 1.5+#2, 7.0+#3) -- (#1, 7.3+#2, 4.3+#3);}{}
  \ifthenelse{#5=0}{\draw [black, ultra thick] (#1, 1.5+#2, 7.0+#3) -- (#1, 5.3+#2, 7.3+#3);}{}
  \ifthenelse{#4=0}{\draw [black, ultra thick] (#1, 1.5+#2, 7.0+#3) -- (#1, 1.3+#2, 1.3+#3);}{}
  \ifthenelse{#5=0}{\draw [black, ultra thick] (#1, 7.3+#2, 4.3+#3) -- (#1, 5.3+#2, 7.3+#3);}{}
  \ifthenelse{#4=0}{\draw [black, ultra thick] (#1, 6.8+#2, 1.1+#3) -- (#1, 1.3+#2, 1.3+#3);}{}
\begin{scope}[canvas is yz plane at x=#1]
\fill[draw=black, fill=\fillcolorA] (1.3+#2, 1.3+#3) circle (1);
\ifthenelse{#4=1}{\node at (1.3+#2, 1.3+#3) [black,transform shape, scale=5, xscale=-1] {?};}{}
\fill[draw=black, fill=red] (4.3+#2, 4.3+#3) circle (1);
\ifthenelse{#5=1}{\node at (7.3+#2, 4.3+#3) [black,transform shape, scale=5, xscale=-1] {?};}{}
\fill[draw=black, fill=\fillcolorB] (7.3+#2, 4.3+#3) circle (1);
\ifthenelse{#5=1}{\node at (7.3+#2, 4.3+#3) [black,transform shape, scale=5, xscale=-1] {?};}{}
\fill[draw=black, fill=\fillcolorB] (5.3+#2, 7.3+#3) circle (1);
\ifthenelse{#5=1}{\node at (5.3+#2, 7.3+#3) [black,transform shape, scale=5, xscale=-1] {?};}{}
\fill[draw=black, fill=red] (1.5+#2, 7.0+#3) circle (1);
\fill[draw=black, fill=\fillcolorA] (6.8+#2, 1.1+#3) circle (1);
\ifthenelse{#4=1}{\node at (6.8+#2, 1.1+#3) [black,transform shape, scale=5, xscale=-1] {?};}{}
\end{scope}
}

\newcommand{\simplecube}[7]
{
  \begin{scope}[shift={#1}]
    \fill[white  ,canvas is xy plane at z=#4] (0,0) rectangle (#2,#3);
    \fill[gray!40,canvas is yz plane at x=#2] (0,0) rectangle (#3,#4);
    \fill[gray!10,canvas is xz plane at y=#3] (0,0) rectangle (#2,#4);
    \foreach\i/\j in {0/1, 1/1, 1/0}
    {
      \draw[line#5] (0,#3*\i,#4*\j) --++ (#2,0,0);
      \draw[line#6] (#2*\i,0,#4*\j) --++ (0,#3,0);
      \draw[line#7] (#2*\i,#3*\j,0) --++ (0,0,#4);
    }
  \end{scope}
}

\newcommand{\cubeab}[4]
{
  \begin{scope}[shift={#1}]
    \fill[draw=black, fill=none, canvas is yz plane at x=0] (0,0) rectangle (#2+#3+#4, #2+#3+#4);
    \fill[draw=black, fill=none, canvas is xz plane at y=0] (0,0) rectangle (#2+#3+#4, #2+#3+#4);
    \simplecube{(#2    ,0    ,0    )}{#2}{#2}{#2}{a}{a}{a}
    \simplecube{(#2+#4,#2    ,#2    )}{#3}{#2}{#2}{b}{a}{a}
    \simplecube{(0    ,#2+#4,#2    )}{#2}{#3}{#2}{a}{b}{a}
    \simplecube{(#4,#2+#4,0    )}{#3}{#3}{#2}{b}{b}{a}
    \simplecube{(0    ,0    ,#2+#4)}{#2}{#2}{#3}{a}{a}{b}
    \simplecube{(#2+#4,0    ,#2+#4)}{#3}{#2}{#3}{b}{a}{b}
    \fill[draw=black, fill=none, canvas is xy plane at z=#2+#3+#4] (0,0) rectangle (#2+#3+#4, #2+#3+#4);
    \fill[draw=black, fill=none, canvas is yz plane at x=#2+#3+#4] (0,0) rectangle (#2+#3+#4, #2+#3+#4);
    \fill[draw=black, fill=none, canvas is xz plane at y=#2+#3+#4] (0,0) rectangle (#2+#3+#4, #2+#3+#4);
  \end{scope}
}

\newcommand{\repourl}{https://github.com/iainrolland/GraphProp}

\begin{document}

    \title{Beyond Low Rank: A Graph-Based Propagation Approach to Tensor Completion for Multi-Acquisition Scenarios}

    \author{Iain Rolland\orcidlink{0000-0002-4137-5605}, Sivasakthy Selvakumaran\orcidlink{0000-0002-8591-0702}, and Andrea Marinoni\orcidlink{0000-0001-6789-0915}
    \thanks{
        This work was funded in part by the UK Engineering and Physical Sciences Research Council (EPSRC) [grant number EP/T517847/1];
        the Centre for Integrated Remote Sensing and Forecasting for Arctic Operations (CIRFA) and the Research Council of Norway [RCN grant number 237906];
        the NATALIE project funded by the European Union Horizon Europe Climate program under grant agreement nr. 101112859;
        the Isaac Newton Trust;
        and Newnham College, Cambridge, United Kingdom.
    }
    \thanks{
        Iain Rolland and Sivasakthy Selvakumaran are with the University of Cambridge, Cambridge, United Kingdom (email: imr27@cam.ac.uk; ss683@cam.ac.uk).}
    \thanks{
        Andrea Marinoni is with UiT the Arctic University of Norway, Tromsø, Norway (email: andrea.marinoni@uit.no).
    }
    }
    \date{}

    \maketitle

    \begin{abstract}
      Tensor completion refers to the problem of recovering the missing, corrupted or unobserved entries in data represented by tensors.
      In this paper, we tackle the tensor completion problem in the scenario in which multiple tensor acquisitions are available and do so without placing constraints on the underlying tensor's rank.
      Whereas previous tensor completion work primarily focuses on low-rank completion methods, we propose a novel graph-based diffusion approach to the problem.
        Referred to as GraphProp, the method propagates observed entries around a graph-based representation of the tensor in order to recover the missing entries.
        A series of experiments have been performed to validate the presented approach, including a synthetically-generated tensor recovery experiment which shows that the method can be used to recover both low and high rank tensor entries.
        The successful tensor completion capabilities of the approach are also demonstrated on a real-world completion problem from the field of multispectral remote sensing completion. 
        Using data acquired from the Landsat 7 platform, we synthetically obscure image sections in order to simulate the scenario in which image acquisitions overlap only partially.
        In these tests, we benchmark against alternative tensor completion approaches as well as existing graph signal recovery methods, demonstrating the superior reconstruction performance of our method versus the state of the art.

    \end{abstract}


    \section{Introduction}\label{sec:introduction}

    \IEEEPARstart{T}{ensor} completion presents itself as a problem within many application scenarios, including image reconstruction~\cite{liu2012tensor}, brain signal processing~\cite{zhang2016eeg, shi2015lrtv} and remote sensing applications~\cite{ng2017adaptive, he2019tvtr}.
    Although the task of recovering missing tensor entries appears at first glance as an ill-posed problem, if we can make assumptions the existence of relationships between tensor entries, the redundancies in the problem can be exploited to make the entry recovery possible.

    The assumption which prevails in the existing literature is that the underlying tensor is of low rank ~\cite{he2019tvtr, chen2019nonlocal, fan2017lrtr} and this assumption is used as the basis for their completion algorithms.
    Low-rank completion refers to the family of methods which generalise low-rank matrix completion to higher-order tensors.
    These methods assume that the dataset to be reconstructed (either in matrix or tensor form) would show low variability across the records.
    Although real-world data is rarely perfectly low rank, there are many cases which present themself to exhibit an approximately low-rank property.
    Under this low-rank assumption, a number of algorithms have been presented~\cite{ng2017adaptive, zhao2016tensorring, madathil2018twist} which complete the matrix or tensor using the values that provide the lowest-rank solution satisfying the constraints on the observed entries.

    While low-rank completion methods perform impressively for low-rank data which is missing at random, it cannot perform completion of datasets consisting of strong variability, i.e. high-rank data, or when large contiguous regions are missing.
    An extreme case of the latter being that completion is not possible when any column or row is missing in its entirety~\cite{candes2010matrix}.
    Although this represents an extreme case, it is not uncommon for the missing patterns in real-world completion problems to form large contiguous patches which approach this scenario.
    The majority of existing theoretical work investigating the extent to which tensors are recoverable from partial observations are based on the assumption that observations are made uniformly at random and slowly varying.
    This, in reality, represents only a tiny fraction of tensor completion applications.
    Many applications, for example sensor failures~\cite{zeng2013recovering, wang2006modis} and cloud removal~\cite{meraner2020cloud_removal, singh2018cloudgan, ebel2021cloudgan}, feature missing patterns which could not be considered uniformly random.

    In order to fulfil a need for a tensor completion method which can handle large contiguous sections of missing entries as well as perform completion without placing constraints on the rank of the underlying tensor, we propose a graph-based diffusion approach to tensor completion.
    Rather than making any assumption about the rank of the underlying tensor we consider the scenario when multiple acquisitions of the tensor are available and assume the existence of a relationship which holds across the tensor acquisitions, preserving the nearest neighbourhood relationships of fibers in one of the tensor dimensions.
    This allows a graph-based representation to be constructed by considering the observed similarities exhibited in each of the partially observed tensor acquisitions.
    The graph constructed represents a unifying graph-based representation across acquisitions.
    Using the unifying graph-based tensor representation as the structure upon which propagation of observed entries is performed, we obtain the proposed graph-based propagation approach to tensor completion, named GraphProp.

    For this method, we derive an upper bound on the completion error and relate this to an existing upper bound derived for graph signal recovery in order to validate our approach.
    Experimentally we show that the propagation-based approach is a valid graph signal recovery method using an existing political blog classification dataset, then prove that the GraphProp approach can handle tensor completion without placing constraints on the underlying tensor's rank using a synthetically-generated dataset.
    Finally, we present a real-world analysis using a synthetically obscured remote sensing dataset and benchmark performance against state of the art low-rank completion and graph signal recovery methods.
    In our analysis we show that the proposed method, referred to as GraphProp, can outperform state of the art methods and strategies both in quality of completion and in computational efficiency.

    The rest of the paper is organised as follows.
    Section~\ref{sec:related-work} introduces and reviews the existing literature in tensor completion.
    Section~\ref{sec:methods} gives details of our proposed approach and presents the GraphProp algorithm.
    Section~\ref{subsec:theoretical_results} derives an upper bound on the completion error and relates this to an existing upper bound in literature for graph-based signal recovery.
    Section~\ref{sec:results_and_analysis} presents a series of experiments and makes comparisons to the state of the art methods in order to validate the proposed approach.
    Finally, section~\ref{sec:conclusion} provides an overview of the contribution and concludes with potential directions of future research.

    \section{Related work}\label{sec:related-work}
    In this section, a review of existing methods for tensor completion is presented, with a summary of the strengths and weaknesses of each approach.
    \subsection{Low-rank completion methods}\label{subsec:low-rank-completion}
    The ill-posed problem of recovering a matrix which is only partially observed was shown by~\cite{candes2009exact} to in fact be possible, under the condition that the original matrix was low rank.
    A matrix with dimensions $m \times n$ is considered low rank if it can be factorised into two matrices with dimensions $m \times r$ and $r \times n$ where $r$ is small compared with $m$ and $n$.
    The matrix obtained via a factorisation of this form consists of columns which lie in a $r$-dimensional subspace of $\mathbb{R}^{m}$ and rows which lie in a $r$-dimensional subspace of $\mathbb{R}^{n}$.
    This reduces the degrees of freedom of the problem and limits the variability in terms of the number of linearly independent rows and columns.

    If the original data is indeed low-rank, there are scenarios which allow the partially observed matrix to be used to recover the original data exactly.
    The family of methods which can be described as low-rank completion algorithms aims to exploit this low-rank property to reconstruct the missing entries from the observations.
    To perform this completion they express an optimisation problem which looks to minimise the rank of the completed matrix subject to equality constraints on the entries which are observed.
    Minimisation of the rank directly is NP-hard, so instead the nuclear norm, which is the convex relaxation of rank, is minimised.
    An algorithm for completion of this type was originally proposed for matrix completion~\cite{cai2010svt} and later generalised to tensors by~\cite{liu2012tensor}.
    There is extensive literature (see e.g.,~\cite{oseledets2011tensor, zhao2016tensorring, madathil2018twist}) focusing on developing better algorithms for low-rank tensor completion which are capable of improved-quality reconstructions and the expansion of the limits of recoverability in terms of minimum required observations.
    The theoretical limits of such approaches have also been been comprehensively studied by investigations such as~\cite{pimentel2016characterization, ashraphijuo2019deterministic, ashraphijuo2019low} which provide lower bounds on completability.

    \subsection{Graph-signal recovery}\label{subsec:graph-signal-recovery}
    Although not generally used to represent and recover graph-based tensor representations, the idea of graph-based signal recovery is not new (see e.g.,~\cite{thanou2016pointclouds, egilmez2014spectralanomaly, huang2018gsp, narang2013graph}).
    Graph-signal recovery can be thought of as a subset of the wider and well-studied problem of signal recover in the field of signal processing.
    In any problem which involves signal recovery relating to data which can be considered to lie on a graph, the problem can be referred to as graph signal recovery.
    In~\cite{chen2014signal}, for example, they propose the use of the graph shift operator to minimise total variation in order to perform graph signal inpainting.
    This idea is extended in~\cite{chen2015signal} which generalises the idea across a number of problems relating to recovering signals on graphs, one of which is the signal inpainting problem.

    The assumption which they make in order to perform signal recovery is that the graph signal is smooth as quantified by the measure of graph total variation~\cite{chen2015signal}.
    Expressing the problem as a minimisation of graph total variation with equality constraints on the observed entries, they perform completion by finding the analytical optimum of this minimisation problem.
    The approach assumes knowledge of a graph structure a priori and normalises the adjacency matrix such that the largest eigenvalue is equal to one.
    Naming this approach graph total variation minimisation (GTVM), they proceed to derive an upper bound on the error of the completion.
    Although we draw analogies between our propagation-based approach to completion and GTVM, we do not make the same assumption of smoothness relating to total variation.
    Our quantitative results show that our approach significantly outperforms GTVM in terms of error.

    \section{Methods}\label{sec:methods}
    \subsection{Notation}\label{subsec:notation}
    Scalars, vectors, matrices and tensors are represented using lowercase letters, boldfaced lowercase letters, boldfaced uppercase letters and calligraphic fonts respectively.
    This notation follows the style used by~\cite{kiers2000towards, kolda2009tensor}.
    Using this style of notation, the $i$th element of a vector $\mathbf{x}$ is denoted by $x_{i}$ and an element of a matrix $\mathbf{X}$ is denoted by $x_{ij}$.
    The order or mode of a tensor is defined as the number of dimensions it possesses and by using colon notation all elements in a given dimension may be sliced.
    For example, $\mathbf{x}_{i:}$ represents the $i^{\text{th}}$ row in a matrix $\mathbf{X}$.

    Fibers are vectors obtained by indexing all dimensions of a tensor except one, with a mode-$k$ fiber referring to the fact that the slice is through the $k^{\text{th}}$ dimension, i.e. $\mathbf{x}_{ij:}$ refers to a mode-$3$ fiber of a third-order tensor.

    Slices are matrices obtained by indexing all but two dimensions of a tensor, e.g. $\mathbf{X}_{i::}$.

    \subsection{Graph theory}\label{subsec:graph-theory}
    Consider a graph $\mathcal{G}=\{\mathcal{V}, \mathcal{E}\}$ which is composed of $\vert\mathcal{V}\vert=n$ vertices $v_{1},\ldots,v_{n}$.
    The graph's edges can be represented by an adjacency matrix, $\mathbf{A}\in\mathbb{R}^{n\times n}$, with entries given by
    \begin{equation}
        a_{ij} =
        \begin{cases}
            1, & \text{if}\ (v_{i}, v_{j})\in\mathcal{E}\\
            0, & \text{otherwise} \\
        \end{cases},\label{eq:equation4}
    \end{equation}
    in the unweighted case or if the graph's edges are weighted then $a_{ij}$ is equal to the weight of $(v_{i}, v_{j})\in\mathcal{E}$.
    In this analysis we will consider strictly undirected graphs, which is to say $a_{ij}=a_{ji}$.
    For the avoidance of doubt, we state here that $a_{ii}=0,\forall i\in\{1,\dots,n\}$, which is to say we use the adjacency matrix without self-connections.

    It is also useful to define two further quantities which derive from the adjacency matrix: the degree matrix, $\mathbf{D}$, and the unnormalised graph Laplacian, $\mathbf{L}$.
    The degree matrix is a diagonal matrix in $\mathbb{R}^{n\times n}$ with diagonal entries given by $D_{ii}=\sum_{j} A_{ij}$.
    The unnormalised graph Laplacian is given by $\mathbf{D}-\mathbf{A}$ and therefore is also in $\mathbb{R}^{n\times n}$.

    \subsection{Graph-based tensor representation}\label{subsec:tensors-as-graphs}
\begin{figure}[!t]
    \centering
        \begin{tikzpicture}[3d view={-55}{10},line join=round,scale=0.3]
          \definecolor{GraphPropColour}{rgb}{166, 206, 227}
          \drawcube{4}{9}{9}{0}{0}{5}
          \drawcube{4}{9}{9}{0}{0}{-14}
          \drawcube{4}{9}{9}{8.5}{-13}{5}
          \drawcube{4}{9}{9}{8.5}{-13}{-14}
            \simplegraph{0}{0.3}{5+0.3}{1}{0}
            \simplegraph{8.5}{-13+.3}{5+0.3}{0}{0}
            \simplegraph{0}{.3}{-14+0.3}{0}{1}
            \simplegraph{8.5}{-13+.3}{-14+0.3}{0}{0}
            \begin{scope}[canvas is plane={O(0,0,0)x(0.4819,-0.8762,0)y(0,0,1)}]
              \draw[black, ultra thick, -latex] (.7,5.5) -- (3,0.5) node[right] {};
              \draw[black, ultra thick, -latex] (.7,-5.5) -- (3,-0.5) node[right] {};
              \draw[black, ultra thick, -latex] (.7,0) -- (3,0) node[right] {};
              \draw[black, ultra thick, -latex] (6., -0.5) -- (8.3, -5.5) node[right] {};
              \draw[black, ultra thick, -latex] (6., 0) -- (8.3, 0) node[right] {};
              \draw[black, ultra thick, -latex] (6., 0.5) -- (8.3, 5.5) node[right] {};
              \draw[draw=black, fill=gprop] (3, -7.5) rectangle ++(3,15) node[pos=.5] {\rotatebox{90}{GraphProp}};
              \draw[draw=none, fill=none] (-4, -2) rectangle ++(2,4) node[pos=.5] {\rotatebox{90}{$\boldsymbol{\cdot}\boldsymbol{\cdot}\boldsymbol{\cdot}$}};
              \draw[draw=none, fill=none] (12, -2) rectangle ++(2,4) node[pos=.5] {\rotatebox{90}{$\boldsymbol{\cdot}\boldsymbol{\cdot}\boldsymbol{\cdot}$}};
              \draw[draw=none, fill=none] (-10, 2.5) rectangle ++(3,9) node[pos=.5] {\rotatebox{90}{$\mathcal{H}^{(\Lambda)}$}};
              \draw[draw=none, fill=none] (-10, -16.5) rectangle ++(3,9) node[pos=.5] {\rotatebox{90}{$\mathcal{H}^{(1)}$}};
              \draw[draw=none, fill=none] (5.5, -16.3) rectangle ++(3,9) node[pos=.5] {\rotatebox{90}{$\mathcal{H}^{(1)}$}};
              \draw[draw=none, fill=none] (5.5, 2.7) rectangle ++(3,9) node[pos=.5] {\rotatebox{90}{$\mathcal{H}^{(\Lambda)}$}};
            \end{scope}
        \end{tikzpicture}
        \caption{Graph nodes are used to represent mode$-m$ fibers. Graph edges are used to connect nearby fibers in $\mathbb{R}^{I_{m}}$.
    The observed fibers in each acquisition are used to obtain the graph's adjacency matrix.
  The graph structure is used to propagate observed entries in the inputs (left) such that the missing entries are reconstructed in the outputs (right).}
    \label{fig::tensor_graph}
    \end{figure}
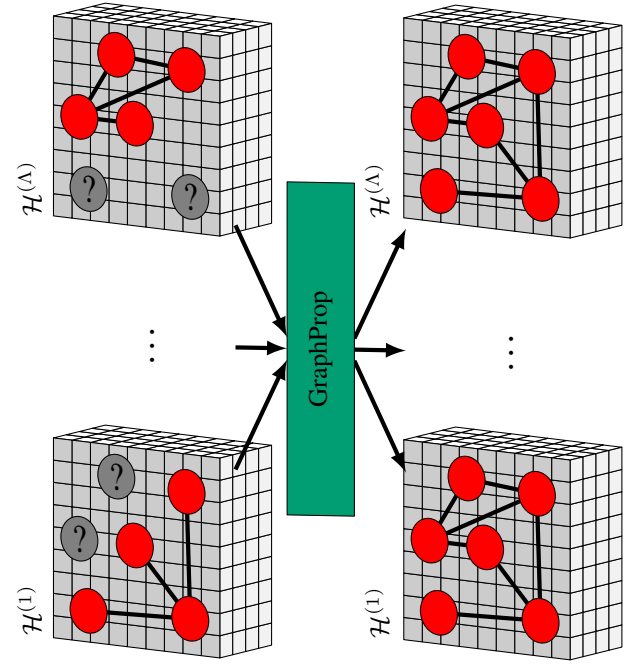

    In this section we describe the graph-based representation of tensors used in the graph-based propagation schema.
    Let us consider a set of $\Lambda$ tensor acquisitions, $\mathcal{H}^{(1)},\dots,\mathcal{H}^{(\Lambda)}$, where $\mathcal{H}^{(\lambda)}\in\mathbb{R}^{I_{1}\times\dots\times I_{m}},\forall\lambda\in\{1,\dots,\Lambda\}$.
    As depicted in figure~\ref{fig::tensor_graph}, the graph representation associates each mode-$m$ fiber with a node in the graph.
    The choice of the $m^{\text{th}}$ mode is arbitrary and will depend on the specifics of the physical meaning of the fibers and the observation patterns of the tensor in the specific application scenario.

    In order to represent each mode-$m$ fiber as a graph node feature, we associate each tensor, $\mathcal{H}^{(\lambda)}$, with a matricisation, $\mathbf{F}^{(\lambda)}$, where $\mathbf{F}^{(\lambda)}=\left(\mathbf{H}^{(\lambda)}_{(m)}\right)^{\top}$.
    As defined in~\cite{kolda2009tensor}, $\mathbf{H}^{(\lambda)}_{(m)}$ refers to the mode-$m$ matricisation of $\mathcal{H}^{(\lambda)}$, which is an operation that places the mode-$m$ fibers in the columns of the matrix.
    By taking the transpose, $\mathbf{F}^{(\lambda)}\in\mathbb{R}^{n\times I_{m}}$, where $n=I_{1}I_{2}\dots I_{m-1}$, is therefore the matrix which contains the mode-$m$ fibers of $\mathcal{H}^{(\lambda)}$ in its rows.
    This provides the feature matrix for the graph corresponding to each tensor acquisition, with the graphs consisting of $n=I_{1}I_{2}\dots I_{m-1}$ nodes.

    With each tensor acquisition we will consider there to exist an observation set, $\Omega^{(\lambda)}\subseteq\{1,\dots,n\}$.
    If $p\in\Omega^{(\lambda)}$, the corresponding fiber, $\mathbf{f}^{(\lambda)}_{p:}$, is considered to be observed in full.
    Conversely, if $p\in\Omega^{(\lambda)}_{c}$, where $\Omega^{(\lambda)}_{c}=\{1,\dots,n\}\backslash\Omega^{(\lambda)}$, the corresponding fiber is considered to be fully unobserved.
    The observation sets are used as indices such that $\mathbf{F}^{(\lambda)}_{\Omega^{(\lambda)}}$ refers to the matrix containing only the rows of $\mathbf{F}^{(\lambda)}$ corresponding to the observed mode-$m$ fibers in the $\lambda^{\text{th}}$ acquisition.

    For each acquisition, an edge set, $\mathcal{E}^{(\lambda)}$ is obtained.
    To do so, each of the observed mode-$m$ fibers is considered as a point in $\mathbb{R}^{I_{m}}$.
    Edges are then obtained by connecting fibers that fall within the $k$-nearest neighbours of another fiber.
    Given that only the observed fibers in each acquisition can be connected, this does not provide edges when $p\in\Omega^{(\lambda)}_{c}$.

    A unifying graph-based representation is obtained by considering the union of the edge sets, i.e. $\mathcal{E}=\left\{\mathcal{E}^{(1)}\cup\dots\cup\mathcal{E}^{(\Lambda)}\right\}$.
    Provided that the observation condition $p\in\left\{\Omega^{(1)}\cup\dots\cup\Omega^{(\Lambda)}\right\},\forall p\in\{1,\dots,n\}$ is met and that $\vert\Omega^{(\lambda)}\vert\geq k+1,\forall \lambda$, every mode-$m$ fiber will have at least $k$ edges in the unifying graph.

    \subsection{Graph-based propagation}\label{subsec:graph-based-diffusion}

    Motivated by the fact that a graph constructed according to section \ref{subsec:tensors-as-graphs} encompasses similarities within the tensor, we suggest that a graph of this kind can act as a structure along which information can be disseminated and therefore we propose graph-based diffusion as an approach to tensor completion.
    This method, represented schematically in figure~\ref{fig::tensor_graph},  aims to exploit the information contained within the observed entries and utilises graph-based propagation to transport this information to the rest of the tensor.

    Diffusion on a graph, as stated by~\cite{kondor2002kernels}, can be performed according to
    \begin{equation}
      \frac{\partial\mathbf{F}}{\partial t}\propto(\mathbf{D}-\mathbf{A})\mathbf{F}=-\mathbf{L}\mathbf{F}.
      \label{eq:diffusion}
    \end{equation}
    Here, $\mathbf{D}$, $\mathbf{A}$ and $\mathbf{L}$ represent the degree, adjacency and unnormalised graph Laplacian as defined in section~\ref{subsec:graph-theory}.
    In this application, they represent the unifying graph obtained by taking the union of edge sets, $\mathcal{E}=\left\{\mathcal{E}^{(1)}\cup\dots\cup\mathcal{E}^{(\Lambda)}\right\}$.
    The $(\lambda)$ superscript has been dropped from the matrix $\mathbf{F}$ for brevity, however, in reality this diffusion is performed for each of the $\Lambda$ acquisitions.

    This derivative is modified to hold the observed entry features as constant in order that the final solution, which is provided by the steady-state solution, matches observed entries in the corresponding locations.
    Consider a reordering of graph nodes such that
    \begin{equation}
  \mathbf{F}\gets  \left[\begin{array}{c}
      \vspace{0.1cm}
      \mathbf{F}_{\Omega} \\
      \mathbf{F}_{\Omega_{c}} \\
  \end{array}\right],
  \label{eq:reodering}
    \end{equation}
without a loss of generality.
This allows the diffusion in~\eqref{eq:diffusion} to incorporate the holding of observed entries as constant when written as
\begin{equation}
  \frac{\partial\mathbf{F}}{\partial t}\propto
  \left[\begin{array}{cc}
      \boldsymbol{0}           & \boldsymbol{0}           \\
      -\mathbf{L}_{\Omega_{c}\Omega} & -\mathbf{L}_{\Omega_{c}\Omega_{c}} \\
  \end{array}\right]
  \mathbf{F},
  \label{eq:steady_state}
\end{equation}
which for the missing fibers gives
\begin{equation}
  \frac{\partial \mathbf{F}_{\Omega_{c}}}{\partial t}\propto-\mathbf{L}_{\Omega_{c}\Omega}\mathbf{F}_{\Omega}-\mathbf{L}_{\Omega_{c}\Omega_{c}}\mathbf{F}_{\Omega_{c}}.
  \label{eq:temp_deriv}
\end{equation}
This propagation can be solved for the steady-state solution either by iteratively applying updates proportional to the derivative or by setting the equation to zero and solving the set of linear equations
\begin{equation}
  \mathbf{L}_{\Omega_{\text{c}}\Omega_{\text{c}}}\mathbf{F}_{\Omega_{\text{c}}}=-\mathbf{L}_{\Omega_{\text{c}}\Omega}\mathbf{F}_{\Omega},
  \label{eq:steady_state_unobserved}
\end{equation}
for $\mathbf{F}_{\Omega_{\text{c}}}$, thus obtaining a completion for the unobserved mode-$m$ fibers.
A summary of the GraphProp approach\footnote{A public repository containing the code used to perform GraphProp has been made available at \repourl.} to tensor completion is provided in algorithm~\ref{alg:cap}.

\begin{algorithm}
\caption{GraphProp algorithm}\label{alg:cap}
\textbf{Input:} $\Omega^{(1)},\dots\Omega^{(\Lambda)}, \mathbf{F}^{(1)}_{\Omega^{(1)}},\dots,\mathbf{F}^{(\Lambda)}_{\Omega^{(\Lambda)}}$
\begin{algorithmic}[1]
  \For{each acquisition $\lambda=1,\dots,\Lambda$}
\State $\mathcal{E}^{(\lambda)}\gets\text{kNN}\left(\mathbf{F}^{(\lambda)}_{\Omega^{(\lambda)}}\right)$ \Comment{$k$-nearest neighbours graph}
  \EndFor
  \State $\mathcal{E}= \left\{\mathcal{E}^{(1)}\cup\dots\cup\mathcal{E}^{(\Lambda)}\right\}$\Comment{combine graphs}
\State $\mathbf{L}=\text{Laplacian}(\mathcal{E})$
  \For{each acquisition $\lambda=1,\dots,\Lambda$}
  \State $\mathbf{F}_{\Omega_{\text{c}}}^{(\lambda)}\gets\text{Solve}\left(\mathbf{L}_{\Omega^{(\lambda)}_{\text{c}}\Omega^{(\lambda)}_{\text{c}}}\mathbf{F}^{(\lambda)}_{\Omega^{(\lambda)}_{\text{c}}}=-\mathbf{L}_{\Omega^{(\lambda)}_{\text{c}}\Omega^{(\lambda)}}\mathbf{F}^{(\lambda)}_{\Omega^{(\lambda)}}\right)$
\EndFor
\end{algorithmic}
\textbf{Output:} $\mathbf{F}^{(1)}_{\Omega_{\text{c}}^{(1)}},\dots, \mathbf{F}^{(\Lambda)}_{\Omega_{\text{c}}^{(\Lambda)}}$
\end{algorithm}

    \section{Error upper bound derivation}\label{subsec:theoretical_results}
    \begin{table}
    \begin{center}
    \caption{Reference table for matrix shorthand relationships used.}
    \label{tab:variables}
    \begingroup
    \setlength{\tabcolsep}{6pt} 
    \begin{tabular}{|c|c|c|}
    \hline
    Shorthand & Longhand & Dimensions\\
    \hline
    \addstackgap{$\mathbf{P}$} & \addstackgap{$\left[\begin{array}{cc} \boldsymbol{0} & \boldsymbol{0} \\ \boldsymbol{0} & \mathbf{I} \\ \end{array}\right]\left(\mathbf{I}-\mathbf{D}^{-1}\mathbf{A}\right)$} & \addstackgap{$n\times n$} \\ \hline
    \addstackgap{$\mathbf{Q}$} & \addstackgap{$\left[\begin{array}{cc} \boldsymbol{0} & \boldsymbol{0} \\ \boldsymbol{0} & \mathbf{I} \\ \end{array}\right]\left(\mathbf{I}+\mathbf{D}^{-1}\mathbf{A}\right)$} & \addstackgap{$n\times n$} \\ \hline
    \addstackgap{$\mathbf{U}$} & \addstackgap{$\mathbf{I}+\mathbf{D}_{\Omega_{c}\Omega_{c}}^{-1}\mathbf{A}_{\Omega_{c}\Omega_{c}}$} & \addstackgap{$\vert\Omega_{c}\vert\times \vert\Omega_{c}\vert$} \\ \hline
    \addstackgap{$\mathbf{V}$} & \addstackgap{$\mathbf{I}-\mathbf{D}_{\Omega_{c}\Omega_{c}}^{-1}\mathbf{A}_{\Omega_{c}\Omega_{c}}$} & \addstackgap{$\vert\Omega_{c}\vert\times \vert\Omega_{c}\vert$} \\ \hline
    \addstackgap{$\mathbf{Y}$} & \addstackgap{$\mathbf{D}_{\Omega_{c}\Omega_{c}}^{-1}\mathbf{A}_{\Omega_{c}\Omega}$} & \addstackgap{$\vert\Omega_{c}\vert\times \vert\Omega\vert$} \\ \hline
    \end{tabular}
    \endgroup
    \end{center}
    \end{table}

    In this section we will derive an upper bound on the estimation error for the steady-state solution of the graph diffusion equation in~\eqref{eq:steady_state_unobserved}.
    To do so, we will follow an approach taken in~\cite{chen2014signal, chen2015signal}, where they consider a graph-based optimisation subject to the equality constraints provided by the observed entries.
    We will first, therefore, rephrase our method in terms of an optimisation problem analagous to that shown in~\cite{chen2014signal, chen2015signal}.

    Considering an optimisation with respect to $\mathbf{F}$, we dictate that $\mathbf{F}_{\Omega}=\mathbf{T}_{\Omega}$, where $\mathbf{T}_{\Omega}$ is the measured entries of the observed nodes, in order to ensure the final solution matches the observations.
    The following optimisation
    \begin{mini}
      {\mathbf{F}}{\vert\vert\mathbf{L}_{\Omega_{c}\Omega}\mathbf{F}_{\Omega}+\mathbf{L}_{\Omega_{c}\Omega_{c}}\mathbf{F}_{\Omega_{c}}\vert\vert_F^2}{}{}
        ,
        \addConstraint{\mathbf{F}_{\Omega}=\mathbf{T}_{\Omega}}
    \end{mini}
    can be seen by inspection to possess a closed-form solution matching the result given in ~\eqref{eq:steady_state_unobserved}.

    To allow us to follow the derivation process in~\cite{chen2014signal, chen2015signal} we must rephrase this optimisation problem slightly.
    By again ordering observed nodes to come before unobserved nodes we can consider block matrices which use $\Omega$ or $\Omega_{c}$ to refer to rows and columns depending on whether they correspond to observed or missing nodes respectively.
    The matrices $\mathbf{P}$, $\mathbf{Q}$, $\mathbf{U}$, $\mathbf{V}$ and $\mathbf{W}$, defined in table~\ref{tab:variables}, are used in this section to ease exposition.
    For example, the matrix $\mathbf{P}$ represents a block matrix given by
    \begin{equation}
        \begin{split}
            \mathbf{P}&=
            \left[\begin{array}{cc}
                      \boldsymbol{0} & \boldsymbol{0} \\
                      \boldsymbol{0} & \mathbf{I}     \\
            \end{array}\right]
            (\mathbf{I}-\mathbf{D}^{-1}\mathbf{A})\\
            &=
            \left[\begin{array}{cc}
                      \boldsymbol{0}                                           & \boldsymbol{0}                                                     \\
                      -\mathbf{D}_{\Omega_{c}\Omega_{c}}^{-1}\mathbf{A}_{\Omega_{c}\Omega} & \mathbf{I}-\mathbf{D}_{\Omega_{c}\Omega_{c}}^{-1}\mathbf{A}_{\Omega_{c}\Omega_{c}} \\
            \end{array}\right]\\
            &=
            \left[\begin{array}{cc}
                      \boldsymbol{0}                                           & \boldsymbol{0}                                                     \\
                      -\mathbf{Y} & \mathbf{V} \\
            \end{array}\right].
        \end{split}\label{eq:equation5}
    \end{equation}
    The matrix $\mathbf{D}$ is diagonal and therefore easily inverted.
    While a graph that has any nodes with degree zero would have a singular $\mathbf{D}$, it is clear from a diffusion perspective, due to these nodes being entirely disconnected from others, that propagation of this form would not provide a completion for these nodes' features.
    We therefore would simply exclude any such nodes and therefore guarantee invertibility.

    We can now rephrase the optimisation problem as
    \begin{argmini}
    {\mathbf{F}}{\left\vert\left\vert\mathbf{P}\mathbf{F}\right\vert\right\vert_F^2}{}{\hat{\mathbf{F}}=}
        ,
        \addConstraint{\mathbf{F}_{\Omega}=\mathbf{T}_{\Omega}},
        \label{eq::optimisation}
    \end{argmini}
    which we will show to again have a closed-form solution matching the result given in ~\eqref{eq:steady_state_unobserved}.

    Since we can express the objective function as
    \begin{equation}
        \vert\vert\mathbf{P}\mathbf{F}\vert\vert^{2}_{F}=
        \left\vert\left\vert-\mathbf{Y}\mathbf{F}_{\Omega} + \mathbf{V}\mathbf{F}_{\Omega_{c}}\right\vert\right\vert^{2}_{F}\label{eq:equation6},
    \end{equation}
    we can consider stationary points by considering when the derivative with respect to $\mathbf{F}_{\Omega_{c}}$ is the zero matrix.
    This is expressed as
    \begin{equation}
\frac{\partial\left(\left\vert\left\vert\mathbf{P}\mathbf{F}\right\vert\right\vert_F^2\right)}{\partial\mathbf{F}_{\Omega_{c}}} =2\mathbf{V}^{\top}(-\mathbf{Y}\mathbf{F}_{\Omega} + \mathbf{V}\mathbf{F}_{\Omega_{c}})=\boldsymbol{0}.
    \end{equation}

    Considering the block matrix form of the Laplacian,
    \begin{equation}
        \mathbf{L}=
        \left[\begin{array}{cc}
                  \mathbf{D}_{\Omega\Omega}-\mathbf{A}_{\Omega\Omega} & -\mathbf{A}_{\Omega\Omega_{c}}                            \\
                  -\mathbf{A}_{\Omega_{c}\Omega}                          & \mathbf{D}_{\Omega_{c}\Omega_{c}} - \mathbf{A}_{\Omega_{c}\Omega_{c}} \\
        \end{array}\right],\label{eq:equation9}
    \end{equation}
    we can substitute $\mathbf{A}_{\Omega_{c}\Omega_{c}}$ for $\mathbf{D}_{\Omega_{c}\Omega_{c}}-\mathbf{L}_{\Omega_{c}\Omega_{c}}$ to obtain a second expression for $\mathbf{V}$, given by
\begin{align}
        \begin{split}
    \mathbf{V}&=\mathbf{I}-\mathbf{D}_{\Omega_{c}\Omega_{c}}^{-1}\mathbf{A}_{\Omega_{c}\Omega_{c}}\\
              &=\mathbf{I}-\mathbf{D}_{\Omega_{c}\Omega_{c}}^{-1}\left(\mathbf{D}_{\Omega_{c}\Omega_{c}}-\mathbf{L}_{\Omega_{c}\Omega_{c}}\right)\\
                    &=\mathbf{D}_{\Omega_{c}\Omega_{c}}^{-1}\mathbf{L}_{\Omega_{c}\Omega_{c}}\label{eq:secondV}.
        \end{split}
\end{align}
By making a similar substitution, we can also obtain a second expression for $\mathbf{Y}$.
Substitution of $\mathbf{A}_{\Omega_{c}\Omega}$ for $-\mathbf{L}_{\Omega_{c}\Omega}$ provides
\begin{equation}
      \mathbf{Y}=\mathbf{D}_{\Omega_{c}\Omega_{c}}^{-1}\mathbf{A}_{\Omega_{c}\Omega}
                =-\mathbf{D}_{\Omega_{c}\Omega_{c}}^{-1}\mathbf{L}_{\Omega_{c}\Omega}\label{eq:secondY}.
\end{equation}
    Using these relationships, our optimal $\mathbf{F}_{\Omega_{c}}$, given that we must replace $\mathbf{F}_{\Omega}$ with $\mathbf{T}_{\Omega}$ to satisfy our equality constraint, can therefore be found according to
    \begin{equation}
\label{eq:optimum}
\begin{split}
  2\mathbf{V}^{\top}\left(-\mathbf{Y}\mathbf{T}_{\Omega} + \mathbf{V}\mathbf{F}_{\Omega_{c}}\right)&=\boldsymbol{0}\\
  2\mathbf{V}^{\top}\left(\mathbf{D}_{\Omega_{c}\Omega_{c}}^{-1}\mathbf{L}_{\Omega_{c}\Omega}\mathbf{T}_{\Omega} + \mathbf{D}_{\Omega_{c}\Omega_{c}}^{-1}\mathbf{L}_{\Omega_{c}\Omega_{c}}\mathbf{F}_{\Omega_{c}}\right)&=\boldsymbol{0}\\
  2\mathbf{V}^{\top}\mathbf{D}_{\Omega_{c}\Omega_{c}}^{-1}\left(\mathbf{L}_{\Omega_{c}\Omega}\mathbf{T}_{\Omega} + \mathbf{L}_{\Omega_{c}\Omega_{c}}\mathbf{F}_{\Omega_{c}}\right)&=\boldsymbol{0}.
\end{split}
    \end{equation}
    From inspection, we can say that when $\mathbf{F}_{\Omega_{c}}=-\mathbf{L}_{\Omega_{c}\Omega_{c}}^{-1}\mathbf{L}_{\Omega_{c}\Omega}\mathbf{T}_{\Omega}$ we have a solution, which matches ~\eqref{eq:steady_state_unobserved}.

    Letting $\mathbf{F}^{0}$ be the \textit{true} signal, and $\hat{\mathbf{F}}$ be our estimation of the signal, we use $\mathbf{W}=\mathbf{F}^{0}-\hat{\mathbf{F}}$ to refer to the estimation error.
    We consider our observed entries to be noiseless and therefore $\vert\vert\mathbf{W}_{\Omega}\vert\vert_{F}=\vert\vert\mathbf{T}_{\Omega} - \hat{\mathbf{F}}_{\Omega}\vert\vert_{F}=0$.
    In order to derive an upper bound on the magnitude of entries in $\mathbf{W}_{\Omega_{c}}$, the section of the estimation error corresponding to the unobserved nodes, we will make use of two inequalities.

   The first inequality is an upper bound on $\vert\vert\mathbf{P}\mathbf{W}\vert\vert_{F}$ and is obtained by making use of the triangle inequality and the fact that $\vert\vert \mathbf{P}\hat{\mathbf{F}} \vert\vert_{F}=0$, as shown below.
    \begin{equation}
        \begin{split}
          \left\vert\left\vert\mathbf{P}\hat{\mathbf{F}}\right\vert\right\vert_F&=\vert\vert-\mathbf{Y}\mathbf{T}_{\Omega}- \mathbf{V}\mathbf{L}_{\Omega_{c}\Omega_{c}}^{-1}\mathbf{L}_{\Omega_{c}\Omega}\mathbf{T}_{\Omega}\vert\vert_{F}\\
                                                                                      &=\left\vert\left\vert-\mathbf{Y}\mathbf{T}_{\Omega} - \mathbf{D}_{\Omega_{c}\Omega_{c}}^{-1}\mathbf{L}_{\Omega_{c}\Omega_{c}}\mathbf{L}_{\Omega_{c}\Omega_{c}}^{-1}\mathbf{L}_{\Omega_{c}\Omega}\mathbf{T}_{\Omega}\right\vert\right\vert_{F}\\
            &=\left\vert\left\vert\mathbf{D}_{\Omega_{c}\Omega_{c}}^{-1}\mathbf{L}_{\Omega_{c}\Omega}\mathbf{T}_{\Omega} - \mathbf{D}_{\Omega_{c}\Omega_{c}}^{-1}\mathbf{L}_{\Omega_{c}\Omega}\mathbf{T}_{\Omega}\right\vert\right\vert_{F}\\
            &=0
        \end{split}
        \label{eq:zero_energy}
    \end{equation}
The upper bound on $\vert\vert\mathbf{P}\mathbf{W}\vert\vert_{F}$ is given by
    \begin{equation}
        \begin{split}
            \vert\vert\mathbf{P}\mathbf{W}\vert\vert_{F}&= \vert\vert\mathbf{P}\left(\mathbf{F}^{0}-\hat{\mathbf{F}}\right)\vert\vert_{F}\\
            &\leq\vert\vert\mathbf{P}\mathbf{F}^{0}\vert\vert_{F}+\cancelto{0}{\vert\vert\mathbf{P}\hat{\mathbf{F}}\vert\vert_{F}}\quad\\
            \vert\vert\mathbf{P}\mathbf{W}\vert\vert_{F} &\leq \vert\psi\vert,
        \end{split}\label{eq:puzzle_part_1}
    \end{equation}
    where the parameter $\psi$ represents the cost of the true signal, i.e.
    \begin{equation}
    \vert\vert\mathbf{P}\mathbf{F}^{0}\vert\vert_{F}^{2}=\psi^{2}.
    \end{equation}

    The second inequality is an upper bound on a conjugate form of the objective function and is presented in (\ref{eq:conjugate_bound}).
        \begin{equation}
      \label{eq:conjugate_bound}
        \begin{split}
            \left\vert\left\vert\mathbf{Q}\mathbf{W}\right\vert\right\vert_{F}
            &=\left\vert\left\vert \left[\begin{array}{cc}
                                             \boldsymbol{0}                                          & \boldsymbol{0}                                                     \\
                                             \mathbf{Y} & \mathbf{U} \\
            \end{array}\right]
            \left[\begin{array}{c}
                      \mathbf{W}_{\Omega}\\ \mathbf{W}_{\Omega_{c}}
            \end{array}\right]\right\vert\right\vert_{F}\\
           &= \left\vert\left\vert \mathbf{Y}\mathbf{W}_{\Omega} + \mathbf{U}\mathbf{W}_{\Omega_{c}}\right\vert\right\vert_{F} \\
           &= \left\vert\left\vert \mathbf{U}\mathbf{W}_{\Omega_{c}}\right\vert\right\vert_{F} \\
           &\leq\left\vert\left\vert\mathbf{U}\right\vert\right\vert_{2}\left\vert\left\vert\mathbf{W}_{\Omega_{c}}\right\vert\right\vert_{F}\\
                &\leq\phi\left\vert\left\vert\mathbf{W}_{\Omega_{c}}\right\vert\right\vert_{F}
        \end{split}
    \end{equation}
    This bound makes use of a second parameter,
    \begin{equation}
        \phi=\left\vert\left\vert \mathbf{U}\right\vert\right\vert_{2},\label{eq:equation15}
    \end{equation}
    which is a function of the graph's structure, where $\vert\vert\cdot\vert\vert_{2}$ when applied to a matrix denotes the spectral norm, given by the largest singular value of the matrix.

  Using the fact that
        \begin{equation}
        \label{eq:conjugates_cancel}
        \begin{split}
            \left\vert\left\vert\left(\mathbf{P}+\mathbf{Q}\right)\mathbf{W}\right\vert\right\vert_{F}
            &=2\left\vert\left\vert \left[\begin{array}{cc}
                                              \boldsymbol{0} & \boldsymbol{0} \\
                                              \boldsymbol{0} & \mathbf{I}     \\
            \end{array}\right]\left[\begin{array}{c}
                                        \mathbf{W}_{\Omega} \\
                                        \mathbf{W}_{\Omega_{c}}
            \end{array}\right]
            \right\vert\right\vert_{F}\\
            &=2\left\vert\left\vert \mathbf{W}_{\Omega_{c}}\right\vert\right\vert_{F},\\
        \end{split}
      \end{equation}
we can combine the results of (\ref{eq:puzzle_part_1}) and (\ref{eq:conjugate_bound}) to obtain
      \begin{equation}
        \label{eq:conjugates_bound}
        \begin{split}
          \vert\vert\mathbf{W}_{\Omega_{c}}\vert\vert_{F}&=
          \frac{1}{2}\left\vert\left\vert\left(\mathbf{P}+\mathbf{Q}\right)\mathbf{W}\right\vert\right\vert_{F}\\
                                                                        &\leq\frac{1}{2}\left(\left\vert\left\vert \mathbf{P}\mathbf{W} \right\vert\right\vert_{F}+\left\vert\left\vert\mathbf{Q}\mathbf{W} \right\vert\right\vert_{F} \right)\\
                                                                        \therefore\vert\vert\mathbf{W}_{\Omega_{c}}\vert\vert_{F}&\leq \frac{1}{2}\left(\vert\psi\vert + \phi\vert\vert\mathbf{W}_{\Omega_{c}}\vert\vert_{F}\right).
        \end{split}
      \end{equation}
    If $\phi<2$ then we can rearrange for the final result: the upper bound on the magnitude of the estimation error corresponding to the unobserved nodes, given by
    \begin{equation}
        \vert\vert\mathbf{W}_{\Omega_{c}}\vert\vert_{F}\leq \frac{\vert\psi\vert}{2-\phi}.
        \label{eq:result}
    \end{equation}

    In terms of when we can say $\phi<2$, we must consider $\left\vert\left\vert \mathbf{I}+\mathbf{D}_{\Omega_{c}\Omega_{c}}^{-1}\mathbf{A}_{\Omega_{c}\Omega_{c}} \right\vert\right\vert_{2}$.
    Using the disc theorem for eigenvalues~\cite{gershgorin1931discs}, we can say that the eigenvalues of $\mathbf{D}_{\Omega_{c}\Omega_{c}}^{-1}\mathbf{A}_{\Omega_{c}\Omega_{c}}$ lie in the interval $(-1, 1)$ and therefore the eigenvalues of $\mathbf{I}+\mathbf{D}_{\Omega_{c}\Omega_{c}}^{-1}\mathbf{A}_{\Omega_{c}\Omega_{c}}$ lie in the interval $(0, 2)$.

    \subsubsection{Comparison with GTVM upper bound}

    In~\cite{chen2015signal} they perform the optimisation which in our notation is given by
    \begin{mini}
    {\mathbf{F}}{\left\vert\left\vert\mathbf{F}-\frac{1}{\vert\lambda_{\max}(\mathbf{A})\vert}\mathbf{A}\mathbf{F}\right\vert\right\vert_F^2}{}{}
        ,
        \addConstraint{\mathbf{F}_{\Omega}=\mathbf{T}_{\Omega}},
    \end{mini}
    where $\lambda_{\max}(\mathbf{A})$ refers to the largest eigenvalue of the adjacency matrix, as introduced in section~\ref{subsec:graph-theory}.
    Using $\mathbf{A}'=\frac{1}{\vert\lambda_{\max}(\mathbf{A})\vert}\mathbf{A}$ to denote the normalised adjacency and under the same noiseless measurement assumptions, their upper bound is given by
    \begin{equation}
        \vert\vert\mathbf{W}_{\Omega_{c}}\vert\vert_F\leq\frac{2\vert\eta\vert}{2-q},
        \label{eq:gtvm_result}
    \end{equation}
    where $\eta^2=\left\vert\left\vert\mathbf{F}^0-\mathbf{A}'\mathbf{F}^0\right\vert\right\vert_F^2$ and $q=\left\vert\left\vert\begin{array}{c}\mathbf{A}'_{\Omega\Omega_{c}}\\\mathbf{I}+\mathbf{A}'_{\Omega_{c}\Omega_{c}}\end{array}\right\vert\right\vert_2$.
    Their $q$ lies in the interval $(0, 2)$ and is analogous to the quantity $\phi$ in our bound.

    \section{Experimental results and analysis}\label{sec:results_and_analysis}
    In this section, we perform a validation of the graph-based propagation approach to tensor completion using a series of experiments.
    This validation consists of three components:
    \begin{enumerate*}[label=(\roman*)]
      \item a graph signal recovery task, which justifies the use of propagation as a graph signal recovery approach;
      \item a synthetically-generated tensor completion task, which allows for a comparison to be made against a low-rank approach to completion and to compare performances as a function of tensor rank;
      \item a real-world application, where we consider multispectral remote sensing acquisitions and perform completion in the scenario where the acquisitions overlap only partially.
      \end{enumerate*}

    \subsection{Classification of political blogs}\label{subsec:blogs}
    In addition to demonstrating that the proposed graph-based tensor representation is valid approach for completion, it is also important to consider whether the proposed propagation approach is itself a valid graph signal recovery approach.
    To do so, we consider a graph signal recovery task and compare our method, GraphProp, with existing graph signal recovery methods in literature.
  
    We consider a blog classification task as performed by~\cite{chen2014signal}, where they present a total variation minimisation approach (GTVM) to graph signal recovery, as well as a total variation regularisation (GTVR) method.
    The dataset consists of $n=1224$ political blogs which was first presented by~\cite{adamic2005political}.
    This experiment acts as a test of the propagation approach only as opposed to the graph-based tensor representation method presented in section~\ref{subsec:tensors-as-graphs} and therefore the graph structure used is as provided in the original dataset, with edges connecting any two blogs which link to each other.

    GraphProp is applied to this problem by taking the observed labels, 0 indicating a left or liberal blog and 1 indicating a right or conservative blog, as node features and propagates holding the observed features as constant until the steady-state is found.
    In our approach, the unlabelled nodes which have a solution value above or below the median are classified as conservative or liberal respectively.
    Accuracy is measured as the percentage of unlabelled nodes which are correctly classified according to the labels in the dataset.
    The observed nodes are randomly sampled with analysis performed for $30$ repeats and values averaged.
    The benchmarked GTVR method is tested at a range of hyperparameter settings, where we use $\alpha$ to indicate the parameter as used by~\cite{chen2015signal}.

    \begin{figure}[!t]
    \centering
    \includegraphics[width=\columnwidth]{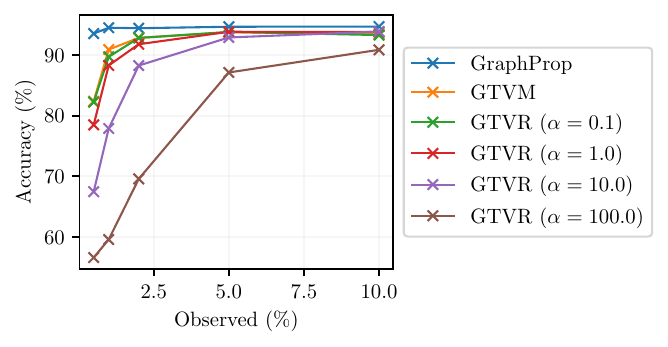}
    \caption{Accuracy of political blog classification as a function of percentage of labels provided.}
    \label{fig::tsp_blog}
    \end{figure}

    The results, shown in figure~\ref{fig::tsp_blog}, show that GraphProp provides improved graph signal recovery performance in terms of classification accuracy versus the GTVM method as well as the GTVR method at all tested hyperparameter settings.
    The biggest improvement is seen at low observation percentages with GraphProp accuracy dropping only slightly whereas benchmarked graph signal recovery methods suffer a steep drop-off in accuracy.
    These results validate the use of graph propagation as a signal recovery method when a signal lies on a graph which consists of edges that connect similar nodes. 

    \subsection{Synthetically-generated dataset}\label{subsec:synth-gen}
    To provide a comparison of the GraphProp approach to tensor completion against the existing approaches relying on low-rank assumptions, an experiment which compares each approach as a function of the underlying tensor's rank has been performed.
    To do so, a synthetically-generated dataset has been created, which allows the Tucker rank to be varied.
    The Tucker decomposition, as introduced in~\cite{tucker1966some}, is given by
\begin{equation}
\mathcal{H}=\mathcal{C}\times_{1}\mathbf{U}_{1}\times_{2}\mathbf{U}_{2}\dots\times_{m}\mathbf{U}_{m},
\end{equation}
where the tensor $\mathcal{H}\in\mathbb{R}^{I_{1}\times\dots\times I_{m}}$ is decomposed into a core tensor, $\mathcal{C}^{r_{1}\times\dots\times r_{m}}$, and a series of unitary matrices, $\mathbf{U}_{1}\in\mathbb{R}^{r_{1}\times I_{1}},\dots,\mathbf{U}_{m}\in\mathbb{R}^{r_{m}\times I_{m}}$.
The $m$-mode product is denoted by $\times_{m}$, and is defined in~\cite{kolda2009tensor}.
The Tucker rank of the tensor $\mathcal{H}$ is given by $(r_{1},\dots,r_{m})$, with $r_{m}=\text{Rank}(U_{m})$.

The first synthetically-generated acquisition tensor, $\mathcal{H}^{(1)}$, was created by randomly generating matrices with orthonormal vectors in the rows.
The entries in the core tensor were randomly sampled from an arbitrary distribution, in this experiment given by $\mathcal{N}(\mu=3,\sigma=3)$.
Subsequent acquisition matricisations were created using a linear transformation of the first tensor's matricisation, described by
\begin{equation}
\mathbf{F}^{(2)}=\mathbf{F}^{(1)}\mathbf{S},
\end{equation}
where $\mathbf{S}$ was a diagonal matrix with entries sampled from $\mathcal{N}(\mu=0,\sigma=1)$.
While this transformation represents a simple scenario with each of the mode-$m$ fiber's channels multiplied by a random scalar between acquisitions, in reality any transformation which locally preserves nearest neighbourhood relationships should suffice.
Note, this does not strictly require a globally linear transformation between acquisitions.

In these results, $\Lambda=2$ observations were generated, with $I_{1}=I_{2}=200$, $I_{3}=3$ and therefore $m=3$.
For each acquisition, an observation set, $\Omega^{(\lambda)}$, was randomly generated with fibers sampled uniformly at random.
To satisfy the observation condition when $\Lambda=2$, i.e. $\left\{\Omega^{(1)}\cup\Omega^{(2)}\right\}=\{1,\dots,n\}$, it is required that if a fiber was missing in the first acquisition that it was observed in the second acquisition.
This limits the maximum missing fiber percentage to $<50\%$ if the same fraction of fibers are missing from each acquisition.

To apply the low-rank completion benchmark the two third-order tensors were stacked in a fourth dimension to provide a tensor of shape $(200, 200, 3, 2)$.
In this study, the High Accuracy Low Rank Tensor Completion (HaLRTC) method of~\cite{liu2012tensor} was applied as the low-rank completion benchmark.

A set of observation tensors of this description were generated to provide a tensor with Tucker rank $(r, r, 3, 2)$, where $r$ is the independent variable which is varied and each method tested for $10$ randomly generated repeats.
Completion accuracy is measured using root mean squared error (RMSE) computed across all missing entries, given by the expression
\begin{equation}
  \text{RMSE}=\sqrt{\frac{\sum_{\lambda=1}^{\Lambda}\sum_{p\in\Omega_{c}^{(\lambda)}}\sum_{i_{m}=1}^{I_{m}}\left(w^{(\lambda)}_{pi_{m}}\right)^{2}}{I_{m}\sum_{\lambda=1}^{\Lambda}\vert\Omega_{c}^{(\lambda)}\vert}},
\end{equation}
where $w^{(\lambda)}_{pi_{m}}$ indexes an entry in the estimation error matrix, $\mathbf{W}^{(\lambda)}$, for the method's completion of the $\lambda^{\text{th}}$ acquisition.

\begin{figure}[!t]
        \centering
        \includegraphics[height=5cm]{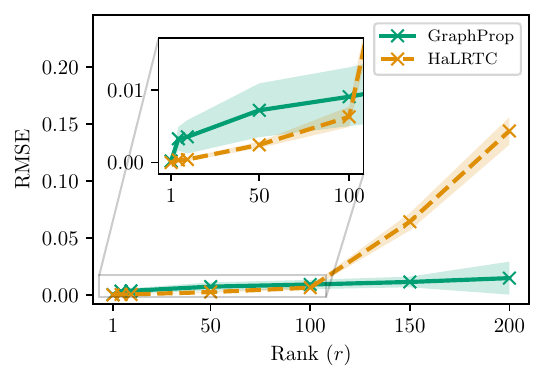}
        \caption{Synthetically-generated tensor completion task with $40\%$ of mode-$3$ fibers missing from each acquisition.
        Results present completion accuracy as measured by RMSE as a function of rank parameter $r$.
          Scenario represents $\Lambda=2$, $m=3$, $I_{1}=I_{2}=200$ and $I_{3}=3$.
        Tucker rank of the stacked acquisition tensor given by $(r, r, 3, 2)$.}
        \label{fig:versus_rank}
    \end{figure}

The results obtained when $40\%$ of fibers are unobserved in each acquisition are presented in figure~\ref{fig:versus_rank}.
Values show the mean across the $10$ repeats and a shaded region representing plus and minus one standard deviation from the mean.
As expected, when the underlying tensor is indeed of low rank, the HaLRTC completion method can effectively exploit the relationships which exist between the tensor entries in order to accurately recover the missing entries.
For ranks of beyond approximately $r=100$, however, the HaLRTC completion accuracy deteriorates rapidly.
Conversely, the GraphProp completion provides comparable but reduced accuracy versus HaLRTC for ranks below $r=100$ but does not exhibit the drop-off in performance as the rank increases.

\begin{figure}[!t]
        \centering
        \includegraphics[height=5cm]{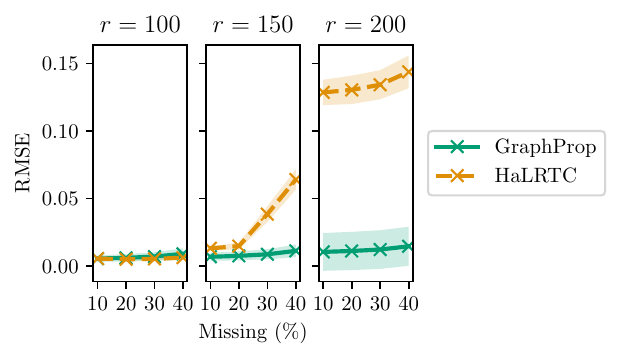}
        \caption{Synthetically-generated tensor completion task with completion accuracy as measured by RMSE as a function of the fraction of fibers missing.
          Tiles show $r=100$, $r=150$ and $r=200$ cases respectively where the Tucker rank of stacked acquisition tensor is given by $(r, r, 3, 2)$.}
        \label{fig:versus_missing}
    \end{figure}

In figure~\ref{fig:versus_missing} the same quantity is presented as a function of the fraction of fibers observed in each acquisition, with a tile depicting when the generated tensor exhibits rank $100$, $150$ and $200$ respectively.
At $r=100$ the two methods perform comparably across observation fractions, while at $r=200$ the GraphProp method significantly outperforms HaLRTC.
At $r=150$ the two methods perform comparably when the percentage of fibers missing is low but as a greater percentage of fibers are considered missing, the low-rank HaLRTC completion performance drops off steeply.

These results indicate that when the local nearest neighbourhood relationships are preserved between the acquisitions that the performance of the GraphProp approach is only weakly conditional on the rank of the underlying tensor.
When this condition holds, GraphProp provides a method capable of completion beyond the domain in which low-rank applications are operative.


    \subsection{Multispectral remote sensing acquisitions}\label{subsec:dataset}
    The final component of our experimentation demonstrates the application of GraphProp to the completion of a real-world tensor completion problem.
    The remote sensing domain is a natural candidate for the multi-acquisition setup of the proposed method as it is generally possible to acquire a time-series of image acquisitions as the satellite platform performs repeat passes over a given region of interest on the ground.
    In this experiment, we consider a multispectral remote sensing completion problem, where missing data in this context might represent one of a number of real-world situations, including sensor failures, cloud obfuscations or partially-overlapping acquisitions.
    Specifically, we have chosen the partially-overlapping scenario as it represents a challenging tensor completion problem and it exists far beyond the low-rank tensor completion assumptions about the entries being observed randomly~\cite{candes2010matrix}.
    The random pattern of observed entries is an assumption upon which low-rank completion algorithms are founded, but it is not always valid in real-world tensor completion applications such as the presented scenario.

    In order to quantitatively benchmark each method, a dataset of remote sensing images which is without missing data was prepared.
    The dataset, which consists of multispectral Landsat 7 acquisitions, covers 50 randomly selected locations across the globe.
    At each location the images acquired by the Landsat 7 platform in the years 2001 and 2002 were downloaded and manually filtered to obtain a cloud-free pair of acquisitions at each location.
    The result is a dataset consisting of 100 images, 2 at each location, each image consisting of 7 spectral bands.
    Spatial dimensions of the images cover 500 pixels in height and width.
    The result is a scenario with $\Lambda=2$ observations of a $m=3$ tensor with $I_{1}=I_{2}=500$ and $I_{3}=7$.

    \begin{figure}[htb]
        \begin{minipage}[b]{.48\linewidth}
          \centering
          \centerline{\includegraphics[width=2.4cm]{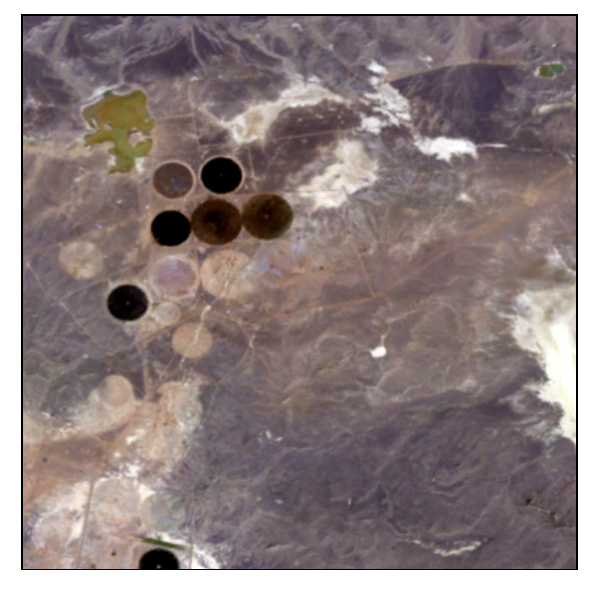}}
          \centerline{(a)}\medskip
        \end{minipage}
        \hfill
        \begin{minipage}[b]{0.48\linewidth}
          \centering
          \centerline{\includegraphics[width=2.4cm]{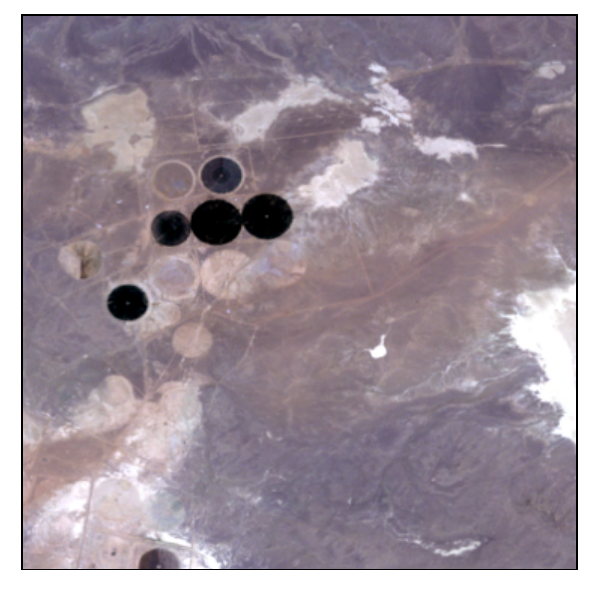}}
          \centerline{(b)}\medskip
        \end{minipage}

        \begin{minipage}[b]{.48\linewidth}
          \centering
          \centerline{\includegraphics[width=2.4cm]{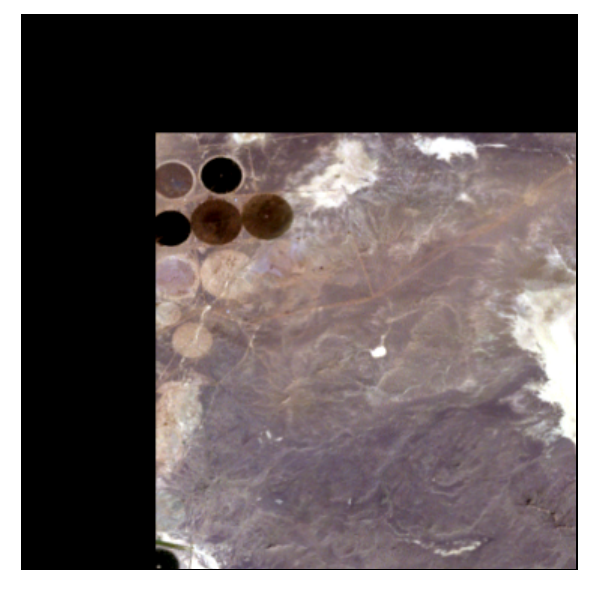}}
          \centerline{(c)}\medskip
        \end{minipage}
        \hfill
        \begin{minipage}[b]{0.48\linewidth}
          \centering
          \centerline{\includegraphics[width=2.4cm]{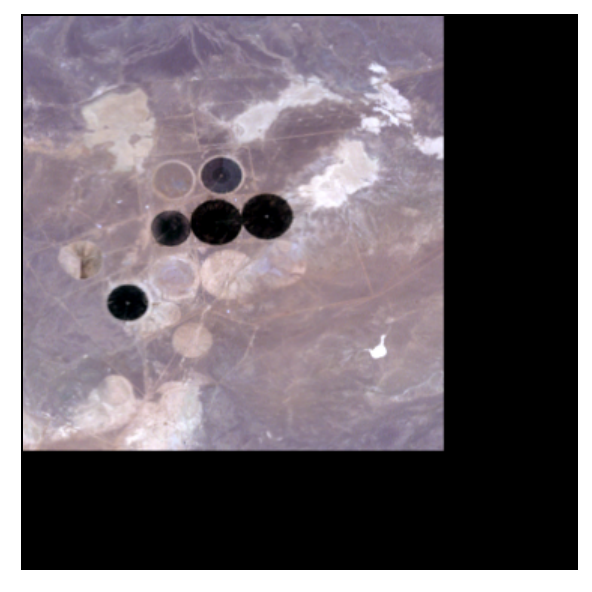}}
          \centerline{(d)}\medskip
        \end{minipage}

        \begin{minipage}[b]{.48\linewidth}
          \centering
          \centerline{\includegraphics[width=2.4cm]{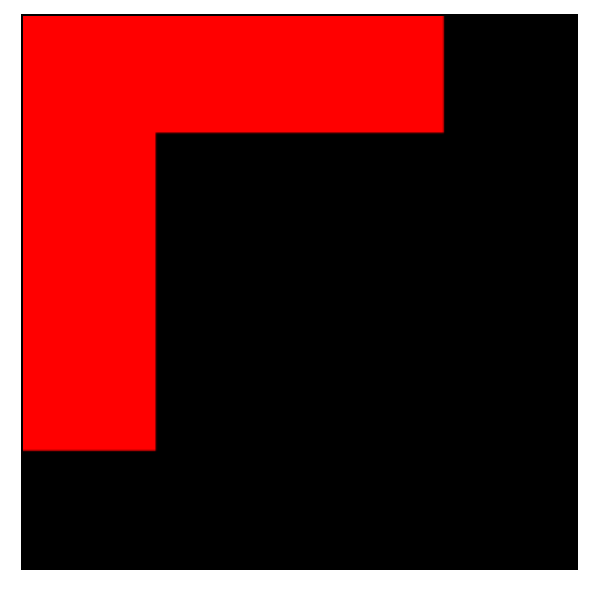}}
          \centerline{(e)}\medskip
        \end{minipage}
        \hfill
        \begin{minipage}[b]{0.48\linewidth}
          \centering
          \centerline{\includegraphics[width=2.4cm]{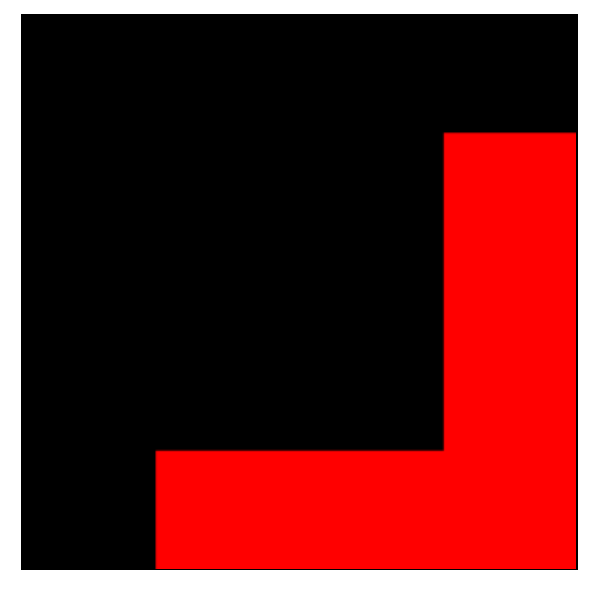}}
          \centerline{(f)}\medskip
        \end{minipage}

        \caption{Simulated partial overlap experiment. The full images are shown in (a) and (b) representing the first and second acquisitions respectively. The images provided representing a partial overlap experiment where 40\% of the original area has been removed are shown in (c) and (d) taken from data in (a) and (b) respectively. The regions which are to be completed and over which metrics are computed are shown for (c) and (d) in red in (e) and (f) respectively.}
        \label{fig:partial_overlap}
    \end{figure}

    To validate our approach we simulate missing data by artificially removing image sections and comparing the quality each method's completion against the original images.
    The image sections are removed so as to simulate a scenario where two acquisitions have been made which overlap only partially.
    This represents the common case where satellite orbits have followed different paths and therefore have provided data which do not perfectly align.
    The goal in this setup is to provide a completed image in the regions which were observed in only one image and therefore lie outwith the region of overlap.
    This problem, to the best of our knowledge, has not been studied previously.
    To simulate this scenario we take each location's Landsat 7 image pair in the aforementioned dataset and remove a fixed number of rows and columns from the left and top of the first-acquired image and the same fixed number of rows and columns from the right and bottom of the second image.
    In testing, the number of rows and columns removed has been varied in order to explore how each method performs as the extent of overlap is varied.
    The amount of data removed is expressed as the area removed as a percentage of the full image area.
    An example of how the partial overlap experiment is simulated is shown in figure~\ref{fig:partial_overlap}.

    \subsection{Experimental results}\label{subsec:experimental_results}
    Benchmarks have been provided to compare our approach to state-of-the-art methods in both low-rank completion as well as graph-based signal recovery.
    From the field of low-rank completion, a comparison is again made to HaLRTC~\cite{liu2012tensor} as well as to an adaptation of HaLRTC, named AWTC~\cite{ng2017adaptive}, which was proposed as a low-rank completion method for the completion of remote sensing data.
    The graph-signal recover benchmark is provided by graph signal inpainting via total variation minimisation (GTVM)~\cite{chen2015signal}.
    The graph-based tensor representation is constructed according to section~\ref{subsec:tensors-as-graphs} using $k=10$ nearest neighbours, with the same graph structure used for both the GTVM benchmark as for the GraphProp implementation.
    The low-rank completion approaches have been applied on the tensor obtained by stacking the two third-order tensors in the fourth dimension giving a tensor of shape $(500, 500, 7, 2)$.

\begin{figure*}[htb]
\begin{minipage}[b]{.23\linewidth}
  \centering
  \centerline{\includegraphics[width=3.6cm]{figures/partial_overlap_1}}
  \centerline{(a)}\medskip
\end{minipage}
\hfill
\begin{minipage}[b]{.23\linewidth}
  \centering
  \centerline{\includegraphics[width=3.6cm]{figures/partial_overlap_2}}
  \centerline{(b)}\medskip
\end{minipage}
\hfill
\begin{minipage}[b]{.23\linewidth}
  \centering
  \centerline{\includegraphics[width=3.6cm]{figures/partial_overlap_1_gt}}
  \centerline{(c)}\medskip
\end{minipage}
\hfill
\begin{minipage}[b]{0.23\linewidth}
  \centering
  \centerline{\includegraphics[width=3.6cm]{figures/partial_overlap_2_gt}}
  \centerline{(d)}\medskip
\end{minipage}

\begin{minipage}[b]{.23\linewidth}
  \centering
  \centerline{\includegraphics[width=3.6cm]{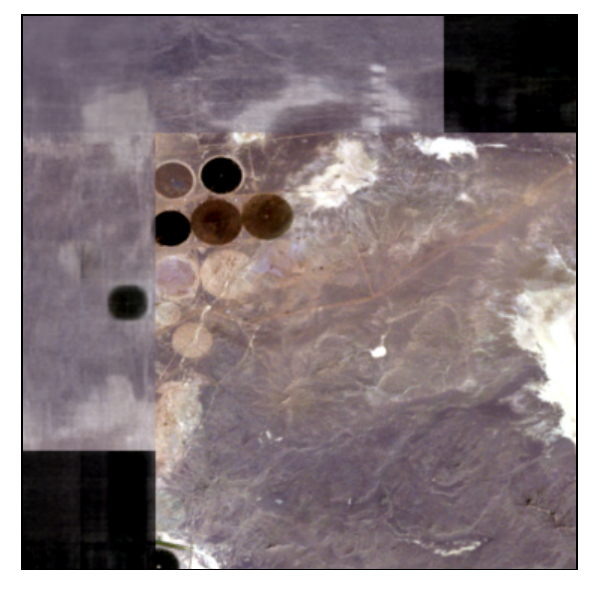}}
  \centerline{(e)}\medskip
\end{minipage}
\hfill
\begin{minipage}[b]{.23\linewidth}
  \centering
  \centerline{\includegraphics[width=3.6cm]{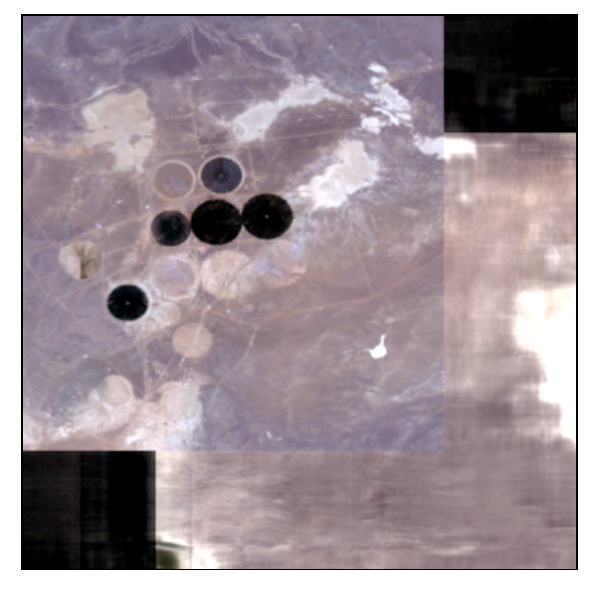}}
  \centerline{(f)}\medskip
\end{minipage}
\hfill
\begin{minipage}[b]{.23\linewidth}
  \centering
  \centerline{\includegraphics[width=3.6cm]{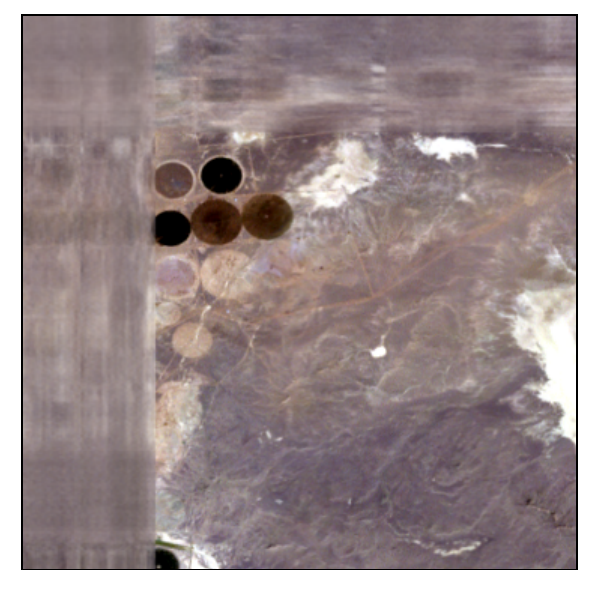}}
  \centerline{(g)}\medskip
\end{minipage}
\hfill
\begin{minipage}[b]{0.23\linewidth}
  \centering
  \centerline{\includegraphics[width=3.6cm]{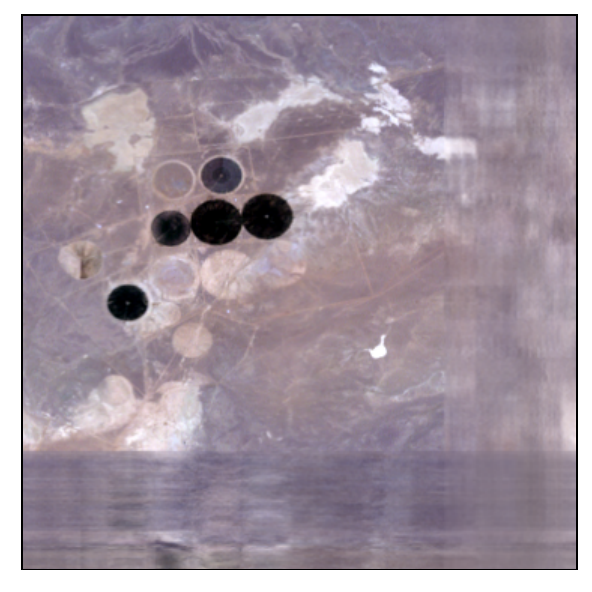}}
  \centerline{(h)}\medskip
\end{minipage}

\begin{minipage}[b]{.23\linewidth}
  \centering
  \centerline{\includegraphics[width=3.6cm]{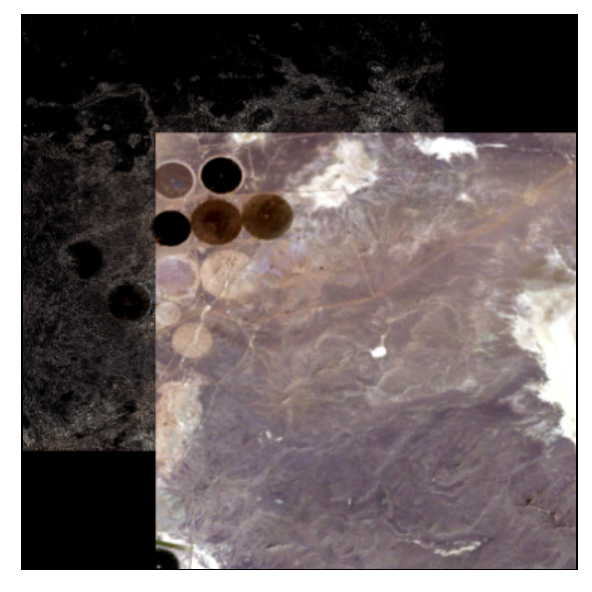}}
  \centerline{(i)}\medskip
\end{minipage}
\hfill
\begin{minipage}[b]{.23\linewidth}
  \centering
  \centerline{\includegraphics[width=3.6cm]{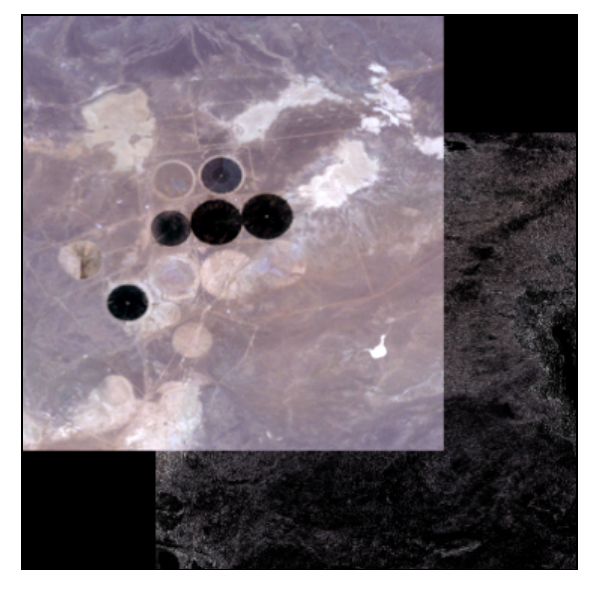}}
  \centerline{(j)}\medskip
\end{minipage}
\hfill
\begin{minipage}[b]{.23\linewidth}
  \centering
  \centerline{\includegraphics[width=3.6cm]{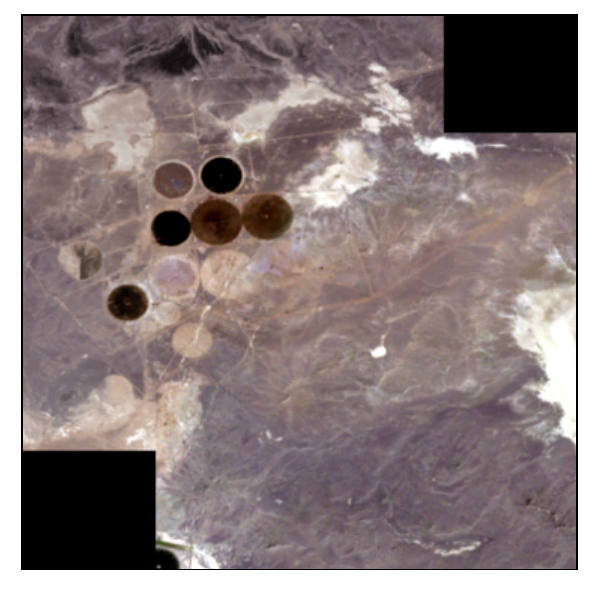}}
  \centerline{(k)}\medskip
\end{minipage}
\hfill
\begin{minipage}[b]{0.23\linewidth}
  \centering
  \centerline{\includegraphics[width=3.6cm]{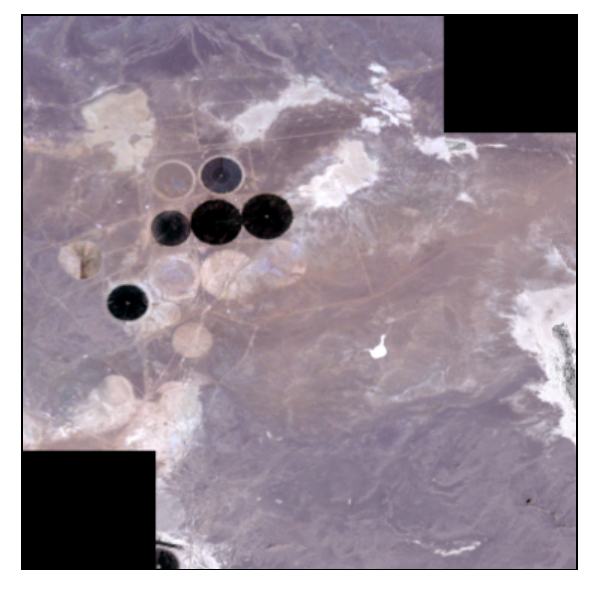}}
  \centerline{(l)}\medskip
\end{minipage}
        \caption{Simulated partial overlap experiment with 40\% area removed. The provided inputs are shown in (a) and (b) representing the first and second acquisitions respectively.
        The ground truth is depicted in (c) and (d). In (e) and (f) the AWTC completion is shown.
        Panels (g) and (h) show the HaLRTC results.
        The GTVM completion is depicted in (i) and (j). In (k) and (l) the GraphProp completion is shown.}
\label{fig:partial_overlap_example}
\end{figure*}

    To quantitatively compare the quality of the recovery we compute four metrics to evaluate the performance of each method.
    These are the mean squared error (MSE), root mean squared error (RMSE), mean absolute error (MAE) and mean peak signal-to-noise ratio (mPSNR) and they are defined as
  \begin{equation}
    \text{MSE}=\frac{1}{I_{m}\sum_{\lambda=1}^{\Lambda}\vert\Omega_{c}^{(\lambda)}\vert}\vert\vert\mathbf{W}_{\Omega_{c}}\vert\vert_{F}^{2}
    \label{eq:MSE}
  \end{equation}
  \begin{equation}
    \text{RMSE}=\frac{1}{I_{m}\sum_{\lambda=1}^{\Lambda}\vert\Omega_{c}^{(\lambda)}\vert}\vert\vert\mathbf{W}_{\Omega_{c}}\vert\vert_{F}
    \label{eq:RMSE}
  \end{equation}
  \begin{equation}
  \text{MAE}=\frac{\sum_{i_{m}=1}^{I_{m}}\left\vert\left\vert\mathbf{w}_{\Omega_{c}i_{m}}\right\vert\right\vert_{1}
}{I_{m}\sum_{\lambda=1}^{\Lambda}\vert\Omega_{c}^{(\lambda)}\vert}    \label{eq:MAE}
  \end{equation}
  \begin{equation}
    \text{mPSNR}=\frac{1}{I_{m}}\sum_{i_{m}=1}^{I_{m}}10\log_{10}\left(\frac{\vert\vert\mathbf{w}_{\Omega_{c}i_{m}}\vert\vert_{\infty}}{\frac{1}{\vert\Omega_{c}\vert}\vert\vert\mathbf{w}_{\Omega_{c}i_{m}}\vert\vert_{2}^{2}}\right),
    \label{eq:mPSNR}
  \end{equation}
  where $\vert\vert\cdot\vert\vert_{\infty}$, $\vert\vert\cdot\vert\vert_{2}$ and $\vert\vert\cdot\vert\vert_{1}$ of a vector denotes the $l_{\infty}$, $l_{2}$ and $l_{1}$ norm respectively.
    Where a pixel was missing from every acquisition these were removed from the analysis which means that the computed metric measures the quality of recovery only for the pixels which were observed at least once.
    The mPSNR metric represents the mean of the PSNR values computed for each band $i_{m}\in\{1,\dots,I_{m}\}$.
  
    \begin{table}
        \begin{center}
            \caption{Simulated partial overlap experiment.
            40\% of area removed.
          Values in red and blue show best and second-best method respectively.}
            \label{tab:partial_overlap}
            \begin{tabular}{| c | c | c | c | c |}
                \hline
                Method & MSE & RMSE & MAE & mPSNR (dB)\\
                \hline
                AWTC & $0.0669$ & $0.2472$ & $0.1917$ & $12.5322$ \\
                \hline
                HaLRTC & $\color{blue}0.0413$ & $\color{blue}0.1962$ & $\color{blue}0.1495$ & $\color{blue}14.4102$ \\
                \hline
                GTVM & $0.1602$ & $0.3963$ & $0.3341$ & $8.1246$ \\
                \hline
                GraphProp & $\mathbf{\color{red}0.0117}$ &$\mathbf{\color{red}0.1020}$ & $\mathbf{\color{red}0.0681}$ & $\mathbf{\color{red}20.3313}$ \\
                \hline
            \end{tabular}
        \end{center}
    \end{table}

    \begin{figure*}[!t]
    \centering
    \includegraphics[width=1.8\columnwidth]{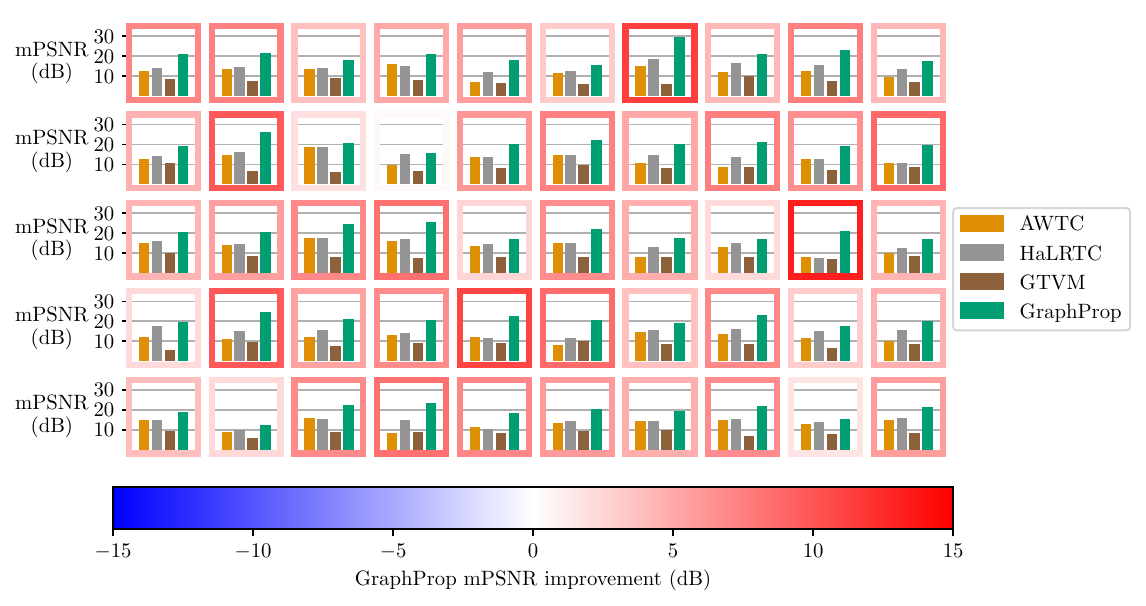}
    \caption{For each of the 50 locations the computed mPSNR for each method's completion in the partial overlap experiment with 40\% area removed.
    The border of each bar chart is coloured according to the horizontal gradient and represents the mPSNR obtained using GraphProp minus the mPSNR of the best benchmarked method.}
    \label{fig:partial_quick_bars}
    \end{figure*}

    The performance of each method at 40\% area removed can be found in table~\ref{tab:partial_overlap}.
    These values represent the values obtained by averaging performance across the 50 locations.
    The values in table~\ref{tab:partial_overlap} show a significantly improved tensor completion is obtained using GraphProp than with even the next-best performance obtained using HaLRTC.
    The improvement is even more significant when compared with either AWTC or GTVM.
    A visual representation of these results is presented in figure~\ref{fig:partial_overlap_example}.
    The two acquisitions differ slightly in terms of their illumination and colouring which allows us to assess each method's robustness to variation between acquisitions.
    The GraphProp method provides results which are visibly superior to those obtained using the benchmarked methods.
    In particular, the GTVM method performs very poorly at this task.
    Of the two low-rank completion methods the HaLRTC algorithm handles this scenario better.
    Both, however, show significantly inferior reconstruction ability for this task when compared against GraphProp.
    For example, AWTC exhibits a tendency to almost directly transfer values across the temporal dimension.
    This provides a reconstruction which suffers from sharp edge artifacts and colour discontinuities.
    The HaLRTC algorithm much better handles the temporal differences with a completion that more closely matches the spectral appearance of the ground truth scene.
    It does, however, suffer from horizontal and vertical banding.
    This is likely a direct consequence of the low-rank assumption, with the optimisation favouring solutions consisting of similar rows and columns.
    It should be pointed out that the HaLRTC is the only method which makes a meaningful attempt at completion of the corners which were observed in neither input.
    As described previously, these regions do not contribute to the values of the computed metrics.

    Dissecting the results into individual locations, displayed in figure~\ref{fig:partial_quick_bars}, demonstrates the consistency of the GraphProp approach, with the method providing the highest mPSNR among the tested methods not only on average but at all of the 50 tested locations.

    \begin{figure}[!t]
        \centering
        \includegraphics[width=\columnwidth]{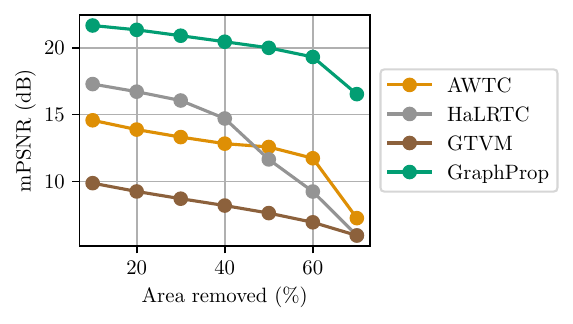}
        \caption{Performance of each method in the partial overlap experiment as the percentage of area removed is varied.
        The plotted series represent the averages obtained across the 50 locations tested.}
        \label{fig:increasingly_partial}
      \end{figure}

\begin{figure*}[htb]
\begin{minipage}[b]{.23\linewidth}
  \centering
  \centerline{\includegraphics[width=3.6cm]{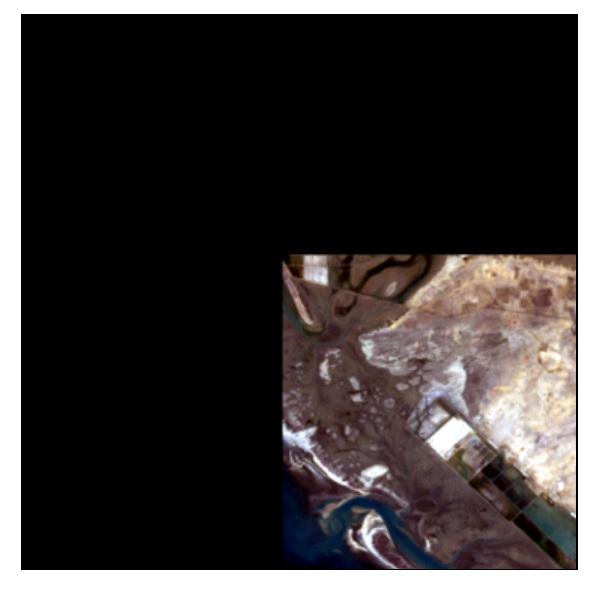}}
  \centerline{(a)}\medskip
\end{minipage}
\hfill
\begin{minipage}[b]{.23\linewidth}
  \centering
  \centerline{\includegraphics[width=3.6cm]{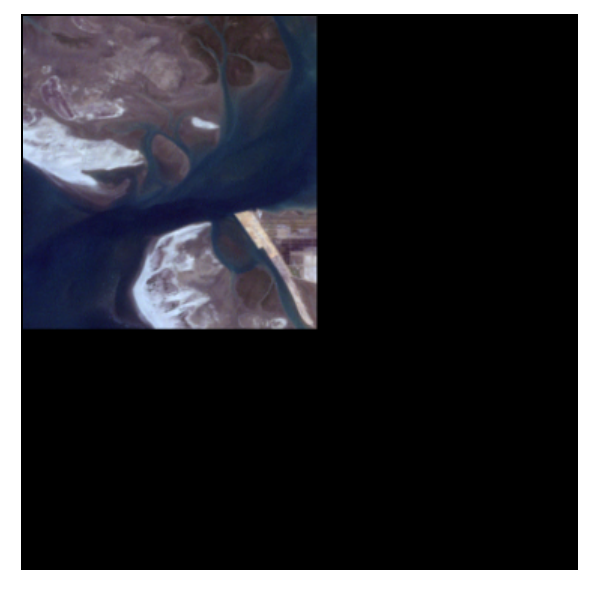}}
  \centerline{(b)}\medskip
\end{minipage}
\hfill
\begin{minipage}[b]{.23\linewidth}
  \centering
  \centerline{\includegraphics[width=3.6cm]{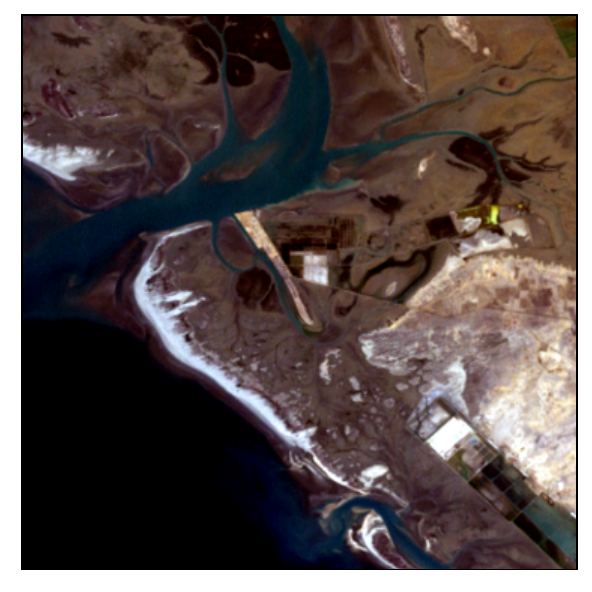}}
  \centerline{(c)}\medskip
\end{minipage}
\hfill
\begin{minipage}[b]{0.23\linewidth}
  \centering
  \centerline{\includegraphics[width=3.6cm]{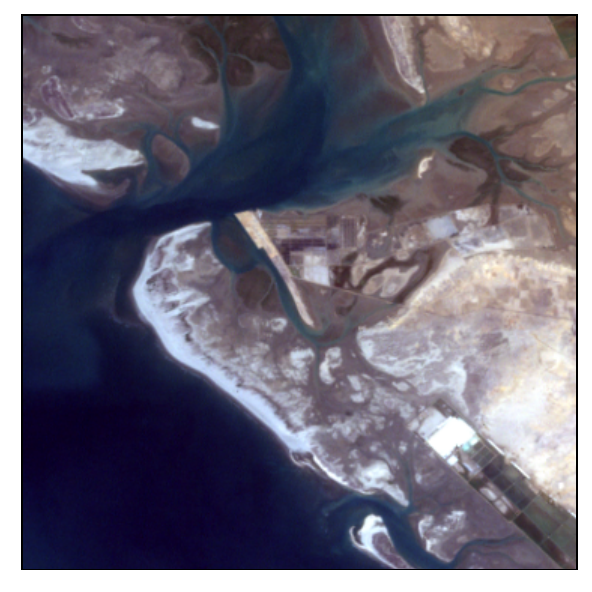}}
  \centerline{(d)}\medskip
\end{minipage}

\begin{minipage}[b]{.23\linewidth}
  \centering
  \centerline{\includegraphics[width=3.6cm]{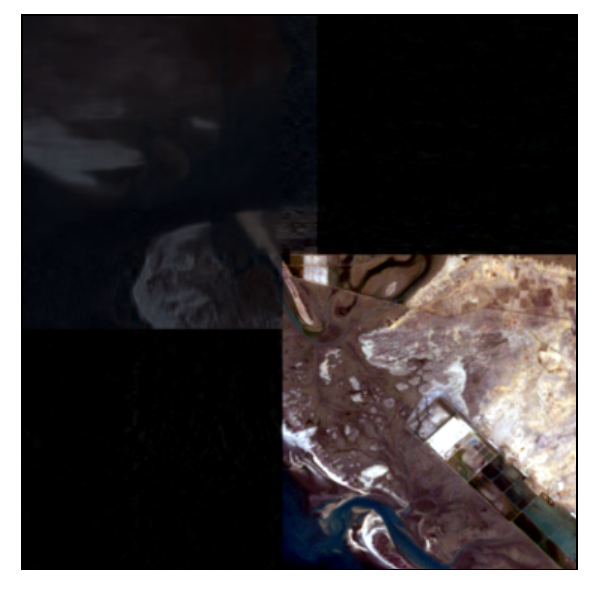}}
  \centerline{(e)}\medskip
\end{minipage}
\hfill
\begin{minipage}[b]{.23\linewidth}
  \centering
  \centerline{\includegraphics[width=3.6cm]{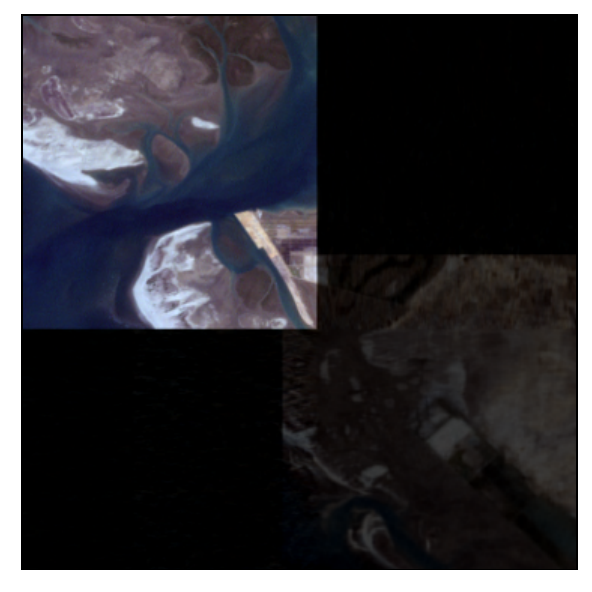}}
  \centerline{(f)}\medskip
\end{minipage}
\hfill
\begin{minipage}[b]{.23\linewidth}
  \centering
  \centerline{\includegraphics[width=3.6cm]{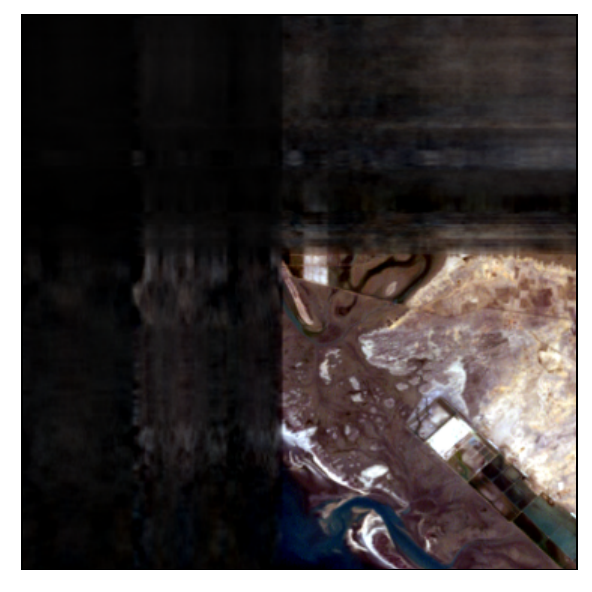}}
  \centerline{(g)}\medskip
\end{minipage}
\hfill
\begin{minipage}[b]{0.23\linewidth}
  \centering
  \centerline{\includegraphics[width=3.6cm]{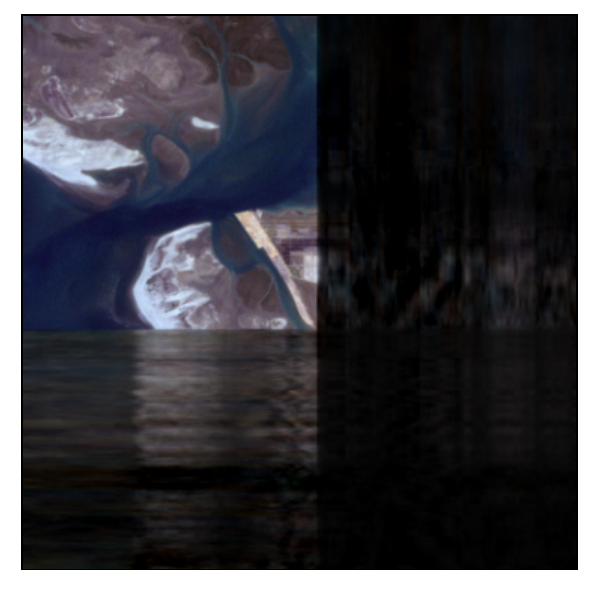}}
  \centerline{(h)}\medskip
\end{minipage}

\begin{minipage}[b]{.23\linewidth}
  \centering
  \centerline{\includegraphics[width=3.6cm]{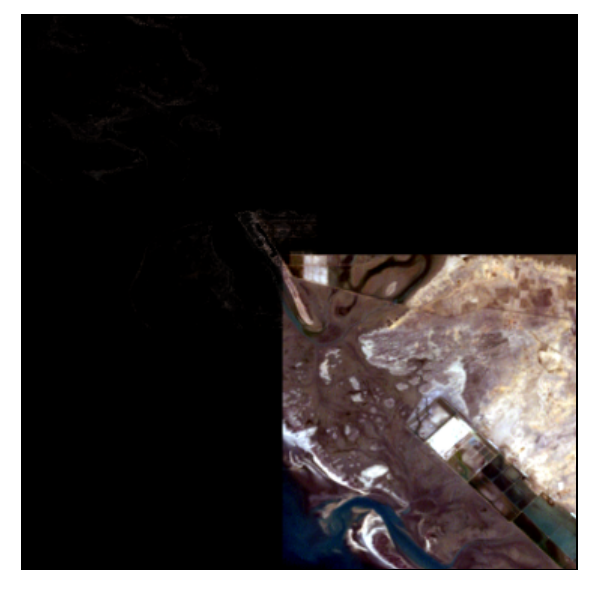}}
  \centerline{(i)}\medskip
\end{minipage}
\hfill
\begin{minipage}[b]{.23\linewidth}
  \centering
  \centerline{\includegraphics[width=3.6cm]{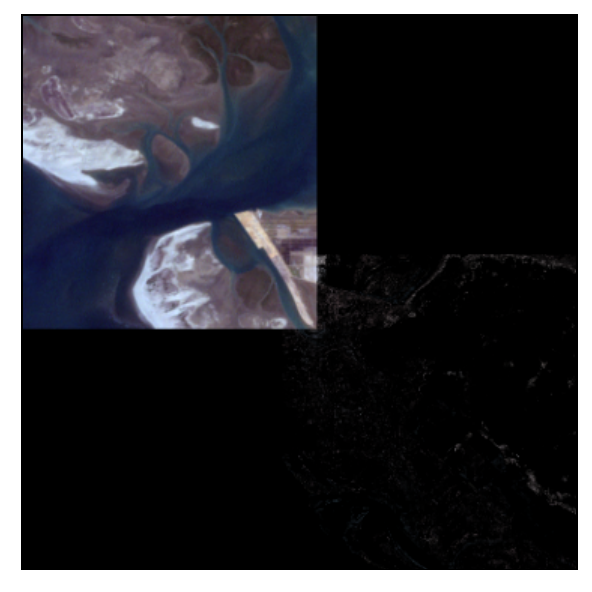}}
  \centerline{(j)}\medskip
\end{minipage}
\hfill
\begin{minipage}[b]{.23\linewidth}
  \centering
  \centerline{\includegraphics[width=3.6cm]{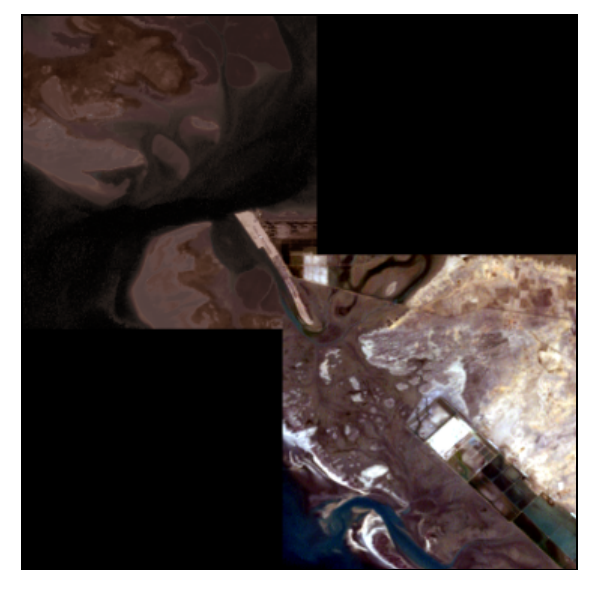}}
  \centerline{(k)}\medskip
\end{minipage}
\hfill
\begin{minipage}[b]{0.23\linewidth}
  \centering
  \centerline{\includegraphics[width=3.6cm]{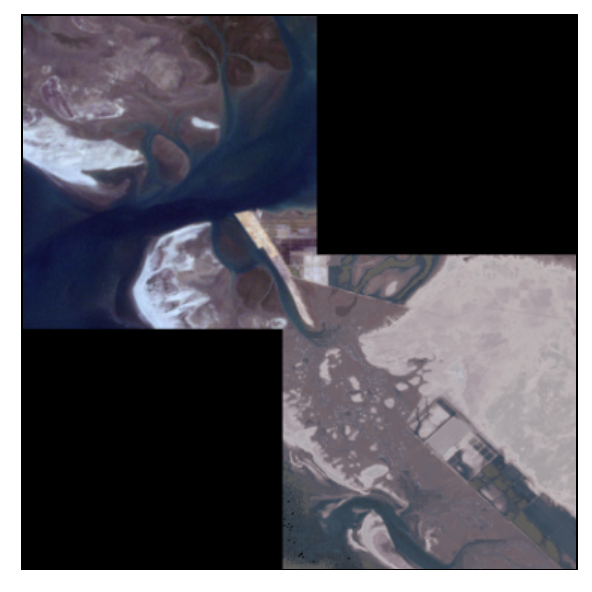}}
  \centerline{(l)}\medskip
\end{minipage}
\caption{Extreme case test of partial overlap where 70\% of both images has been removed leaving only 0.8\% of the spatial area observed in both images. The provided inputs are shown in (a) and (b) representing the first and second acquisitions respectively.
  The ground truth is depicted in (c) and (d). In (e) and (f) the AWTC completion is shown.
  Panels (g) and (h) show the HaLRTC results.
The GTVM completion is depicted in (i) and (j). In (k) and (l) the GraphProp completion is shown.}
\label{fig:partial_overlap_example_70}
\end{figure*}

    As an extension to the partial overlap experiment, an investigation which considers how each method handles varying amounts of overlap between the two images was conducted.
    By varying the amount of area removed between 10\% and 70\% we show performance as a function of overlap extent.
    Tests with beyond 70\% area removed were not conducted as they would represent a scenario in which no overlap between the two images was given.
    The performance as measured by mPSNR for this test is presented in figure~\ref{fig:increasingly_partial}.
    This figure shows that GraphProp achieves a higher mPSNR reconstruction across all scenarios.
    Although there is an accelerated deterioration in mPSNR beyond 60\% area removed, the GraphProp approach still achieves results at 70\% area removed which are approximately comparable with the mPSNR provided by AWTC and HaLRTC at only 10\% area removed.
    A depiction of this performance at 70\% area removed is provided in figure~\ref{fig:partial_overlap_example_70}.
    Despite only providing an overlap of below 1\% of total area, GraphProp can be used to provide a meaningful completion of the missing regions, which none of the benchmark methods are capable of.
    
    \begin{table}
        \begin{center}
            \caption{Computation times (partial overlap experiment with 40\% area removed).}
            \label{tab:computation_times_partial}
            \begin{tabular}{| c | c |}
                \hline
                Method & Average runtime (seconds)\\
                \hline
                AWTC & $261.9$ \\ 
                \hline
                HaLRTC & $221.1$ \\ 
                \hline
                GTVM & $122.4$ \\ 
                \hline
                GraphProp & $142.0$ \\ 
                \hline
            \end{tabular}
        \end{center}
    \end{table}

    In addition to a comparison of the quality of the completions, we also provide a comparison of the computation times of each method.
    In each case the experiment has been performed in Python on a desktop computer with an Intel Core i9-10900X CPU and 32GB of RAM.

    Each method's average runtime for the partial overlap experiment tested with 40\% area removed can be seen in table~\ref{tab:computation_times_partial}.
    The runtime represents the entire process including the time take to compute the $k$-nearest neighbours edges.
    The two graph-based methods are performed in approximately half the time needed to perform the low-rank tensor completion algorithms.

    \section{Conclusion}\label{sec:conclusion}
    In this paper we have presented a novel tensor completion method which applies to multi-acquisition scenarios without constraints on the rank of the underlying tensor.
    The presented approach operates on the assumption that there exists a relationship which holds across the tensor acquisitions, preserving the nearest neighbourhood relationships of fibers in one of the tensor dimensions.
    By exploiting this relationship, a unifying graph-based representation of the tensor can be constructed by building up an edge set using the similarities which are observed to exist in the individual partially-observed tensor acquisitions.
    This unifying graph-based tensor representation provides the structure upon which diffusion can take place, providing the mechanism for the completion of missing entries.

    The proposed method has been proven experimentally to be robust in challenging tensor completion scenarios and through the use of a set of synthetically-generated tensor completion experiments, this performance is shown to hold across both low and high tensor rank contexts.

    Using a multispectral remote sensing dataset, we demonstrate the effectiveness of this method in a real-world context.
    The proposed GraphProp method is shown to successfully recover tensor entries in scenarios well beyond the domain in which low-rank completion methods are viable.
    In addition to the improved performance, we have also shown that the proposed method is computationally efficient and can be performed quicker than low-rank tensor completion algorithms and comparably with other graph signal recovery methods.

    There is scope for further research to consider extended applications of the approach.
    For example, future work might investigate the limits of the preservation of nearest neighbourhood assumption in order to easily assess whether GraphProp method can be successfully applied to the specifics of different tensor completion applications.
    Further research might also consider how the GraphProp approach should be adapted to cases which involve noisy observations of the underlying tensor's entries.

    \bibliographystyle{IEEEtran}
    \bibliography{refs}

\begin{thebibliography}{10}
\providecommand{\url}[1]{#1}
\csname url@samestyle\endcsname
\providecommand{\newblock}{\relax}
\providecommand{\bibinfo}[2]{#2}
\providecommand{\BIBentrySTDinterwordspacing}{\spaceskip=0pt\relax}
\providecommand{\BIBentryALTinterwordstretchfactor}{4}
\providecommand{\BIBentryALTinterwordspacing}{\spaceskip=\fontdimen2\font plus
\BIBentryALTinterwordstretchfactor\fontdimen3\font minus
  \fontdimen4\font\relax}
\providecommand{\BIBforeignlanguage}[2]{{%
\expandafter\ifx\csname l@#1\endcsname\relax
\typeout{** WARNING: IEEEtran.bst: No hyphenation pattern has been}%
\typeout{** loaded for the language `#1'. Using the pattern for}%
\typeout{** the default language instead.}%
\else
\language=\csname l@#1\endcsname
\fi
#2}}
\providecommand{\BIBdecl}{\relax}
\BIBdecl

\bibitem{liu2012tensor}
J.~Liu, P.~Musialski, P.~Wonka, and J.~Ye, ``{Tensor Completion for Estimating
  Missing Values in Visual Data},'' \emph{IEEE Transactions on Pattern Analysis
  and Machine Intelligence}, vol.~35, no.~1, pp. 208--220, 2012.

\bibitem{zhang2016eeg}
Y.~Zhang, Q.~Zhao, G.~Zhou, J.~Jin, X.~Wang, and A.~Cichocki, ``Removal of eeg
  artifacts for bci applications using fully bayesian tensor completion,'' in
  \emph{2016 IEEE International Conference on Acoustics, Speech and Signal
  Processing (ICASSP)}, 2016, pp. 819--823.

\bibitem{shi2015lrtv}
F.~Shi, J.~Cheng, L.~Wang, P.-T. Yap, and D.~Shen, ``Lrtv: Mr image
  super-resolution with low-rank and total variation regularizations,''
  \emph{IEEE Transactions on Medical Imaging}, vol.~34, no.~12, pp. 2459--2466,
  2015.

\bibitem{ng2017adaptive}
M.~K.-P. Ng, Q.~Yuan, L.~Yan, and J.~Sun, ``{An Adaptive Weighted Tensor
  Completion Method for the Recovery of Remote Sensing Images With Missing
  Data},'' \emph{IEEE Transactions on Geoscience and Remote Sensing}, vol.~55,
  no.~6, pp. 3367--3381, 2017.

\bibitem{he2019tvtr}
W.~He, N.~Yokoya, L.~Yuan, and Q.~Zhao, ``Remote {S}ensing {I}mage
  {R}econstruction {U}sing {T}ensor {R}ing {C}ompletion and {T}otal
  {V}ariation,'' \emph{IEEE Transactions on Geoscience and Remote Sensing},
  vol.~57, no.~11, pp. 8998--9009, 2019.

\bibitem{chen2019nonlocal}
Y.~Chen, W.~He, N.~Yokoya, T.-Z. Huang, and X.-L. Zhao, ``Nonlocal
  {T}ensor-{R}ing {D}ecomposition for {H}yperspectral {I}mage {D}enoising,''
  \emph{IEEE Transactions on Geoscience and Remote Sensing}, vol.~58, no.~2,
  pp. 1348--1362, 2019.

\bibitem{fan2017lrtr}
H.~Fan, Y.~Chen, Y.~Guo, H.~Zhang, and G.~Kuang, ``Hyperspectral image
  restoration using low-rank tensor recovery,'' \emph{IEEE Journal of Selected
  Topics in Applied Earth Observations and Remote Sensing}, vol.~10, no.~10,
  pp. 4589--4604, 2017.

\bibitem{zhao2016tensorring}
Q.~Zhao, G.~Zhou, S.~Xie, L.~Zhang, and A.~Cichocki, ``{T}ensor {R}ing
  {D}ecomposition,'' \emph{arXiv preprint arXiv:1606.05535}, 2016.

\bibitem{madathil2018twist}
B.~Madathil and S.~N. George, ``Twist tensor total variation
  regularized-reweighted nuclear norm based tensor completion for video missing
  area recovery,'' \emph{Information Sciences}, vol. 423, pp. 376--397, 2018.

\bibitem{candes2010matrix}
E.~J. Candes and Y.~Plan, ``Matrix completion with noise,'' \emph{Proceedings
  of the IEEE}, vol.~98, no.~6, pp. 925--936, 2010.

\bibitem{zeng2013recovering}
C.~Zeng, H.~Shen, and L.~Zhang, ``Recovering missing pixels for {L}andsat
  {E}{T}{M}+ {S}{L}{C}-off imagery using multi-temporal regression analysis and
  a regularization method,'' \emph{Remote Sensing of Environment}, vol. 131,
  pp. 182--194, 2013.

\bibitem{wang2006modis}
L.~Wang, J.~Qu, X.~Xiong, X.~Hao, Y.~Xie, and N.~Che, ``{A New Method for
  Retrieving Band 6 of Aqua MODIS},'' \emph{IEEE Geoscience and Remote Sensing
  Letters}, vol.~3, no.~2, pp. 267--270, 2006.

\bibitem{meraner2020cloud_removal}
A.~Meraner, P.~Ebel, X.~X. Zhu, and M.~Schmitt, ``{C}loud removal in
  {S}entinel-2 imagery using a deep residual neural network and
  {S}{A}{R}-optical data fusion,'' \emph{ISPRS Journal of Photogrammetry and
  Remote Sensing}, vol. 166, pp. 333--346, 2020.

\bibitem{singh2018cloudgan}
P.~Singh and N.~Komodakis, ``{Cloud-Gan: Cloud Removal for Sentinel-2 Imagery
  Using a Cyclic Consistent Generative Adversarial Networks},'' in
  \emph{{I}{G}{A}{R}{S}{S} 2018 - 2018 {I}{E}{E}{E} {I}nternational
  {G}eoscience and {R}emote {S}ensing {S}ymposium}, 2018, pp. 1772--1775.

\bibitem{ebel2021cloudgan}
P.~Ebel, A.~Meraner, M.~Schmitt, and X.~X. Zhu, ``{M}ultisensor {D}ata {F}usion
  for {C}loud {R}emoval in {G}lobal and {A}ll-{S}eason {S}entinel-2
  {I}magery,'' \emph{IEEE Transactions on Geoscience and Remote Sensing},
  vol.~59, no.~7, pp. 5866--5878, 2021.

\bibitem{candes2009exact}
E.~J. Candès and B.~Recht, ``{Exact Matrix Completion via Convex
  Optimization},'' \emph{Foundations of Computational Mathematics}, vol.~9,
  no.~6, pp. 717--772, 2009.

\bibitem{cai2010svt}
J.-F. Cai, E.~J. Cand{\`e}s, and Z.~Shen, ``{A} {S}ingular {V}alue
  {T}hresholding {A}lgorithm for {M}atrix {C}ompletion,'' \emph{SIAM Journal on
  Optimization}, vol.~20, no.~4, pp. 1956--1982, 2010.

\bibitem{oseledets2011tensor}
I.~V. Oseledets, ``{T}ensor-{T}rain {D}ecomposition,'' \emph{SIAM Journal on
  Scientific Computing}, vol.~33, no.~5, pp. 2295--2317, 2011.

\bibitem{pimentel2016characterization}
D.~L. Pimentel-Alarc{\'o}n, N.~Boston, and R.~D. Nowak, ``{A}
  {C}haracterization of {D}eterministic {S}ampling {P}atterns for {L}ow-{R}ank
  {M}atrix {C}ompletion,'' \emph{IEEE Journal of Selected Topics in Signal
  Processing}, vol.~10, no.~4, pp. 623--636, 2016.

\bibitem{ashraphijuo2019deterministic}
M.~Ashraphijuo, V.~Aggarwal, and X.~Wang, ``{D}eterministic and {P}robabilistic
  {C}onditions for {F}inite {C}ompletability of {L}ow-{T}ucker-{R}ank
  {T}ensor,'' \emph{IEEE Transactions on Information Theory}, vol.~65, no.~9,
  pp. 5380--5400, 2019.

\bibitem{ashraphijuo2019low}
M.~Ashraphijuo, X.~Wang, and J.~Zhang, ``{L}ow-{R}ank {D}ata {C}ompletion
  {W}ith {V}ery {L}ow {S}ampling {R}ate {U}sing {N}ewton's {M}ethod,''
  \emph{IEEE Transactions on Signal Processing}, vol.~67, no.~7, pp.
  1849--1859, 2019.

\bibitem{thanou2016pointclouds}
D.~Thanou, P.~A. Chou, and P.~Frossard, ``{G}raph-{B}ased {C}ompression of
  {D}ynamic {3D} {P}oint {C}loud {S}equences,'' \emph{IEEE Transactions on
  Image Processing}, vol.~25, no.~4, pp. 1765--1778, 2016.

\bibitem{egilmez2014spectralanomaly}
H.~E. Egilmez and A.~Ortega, ``Spectral anomaly detection using graph-based
  filtering for wireless sensor networks,'' in \emph{2014 IEEE International
  Conference on Acoustics, Speech and Signal Processing (ICASSP)}, 2014, pp.
  1085--1089.

\bibitem{huang2018gsp}
W.~Huang, T.~A.~W. Bolton, J.~D. Medaglia, D.~S. Bassett, A.~Ribeiro, and
  D.~Van De~Ville, ``{A} {G}raph {S}ignal {P}rocessing {P}erspective on
  {F}unctional {B}rain {I}maging,'' \emph{Proceedings of the IEEE}, vol. 106,
  no.~5, pp. 868--885, 2018.

\bibitem{narang2013graph}
S.~K. Narang, A.~Gadde, E.~Sanou, and A.~Ortega, ``Localized iterative methods
  for interpolation in graph structured data,'' in \emph{2013 IEEE Global
  Conference on Signal and Information Processing}, 2013, pp. 491--494.

\bibitem{chen2014signal}
S.~Chen, A.~Sandryhaila, G.~Lederman, Z.~Wang, J.~M. Moura, P.~Rizzo,
  J.~Bielak, J.~H. Garrett, and J.~Kova{\v{c}}evic, ``Signal inpainting on
  graphs via total variation minimization,'' in \emph{2014 IEEE International
  Conference on Acoustics, Speech and Signal Processing (ICASSP)}.\hskip 1em
  plus 0.5em minus 0.4em\relax IEEE, 2014, pp. 8267--8271.

\bibitem{chen2015signal}
S.~Chen, A.~Sandryhaila, J.~M. Moura, and J.~Kovacevic, ``{Signal Recovery on
  Graphs: Variation Minimization},'' \emph{IEEE Transactions on Signal
  Processing}, vol.~63, no.~17, pp. 4609--4624, 2015.

\bibitem{kiers2000towards}
H.~A. Kiers, ``Towards a standardized notation and terminology in multiway
  analysis,'' \emph{Journal of Chemometrics: A Journal of the Chemometrics
  Society}, vol.~14, no.~3, pp. 105--122, 2000.

\bibitem{kolda2009tensor}
T.~G. Kolda and B.~W. Bader, ``Tensor decompositions and applications,''
  \emph{SIAM review}, vol.~51, no.~3, pp. 455--500, 2009.

\bibitem{kondor2002kernels}
{Kondor, Risi and Lafferty, John}, ``Diffusion kernels on graphs and other
  discrete input spaces,'' \emph{ICML}, vol. Vol. 2, 05 2002.

\bibitem{gershgorin1931discs}
S.~A. Geršgorin, ``{Über die Abgrenzung der Eigenwerte einer Matrix},''
  \emph{lzv. Akad. Nauk. USSR. Otd. Fiz-Mat. Nauk}, vol.~7, pp. 749--754, 1931.

\bibitem{adamic2005political}
\BIBentryALTinterwordspacing
L.~A. Adamic and N.~Glance, ``The political blogosphere and the 2004 u.s.
  election: Divided they blog,'' in \emph{Proceedings of the 3rd International
  Workshop on Link Discovery}, ser. LinkKDD '05.\hskip 1em plus 0.5em minus
  0.4em\relax New York, NY, USA: Association for Computing Machinery, 2005, p.
  36–43. [Online]. Available: \url{https://doi.org/10.1145/1134271.1134277}
\BIBentrySTDinterwordspacing

\bibitem{tucker1966some}
L.~R. Tucker, ``Some mathematical notes on three-mode factor analysis,''
  \emph{Psychometrika}, vol.~31, no.~3, pp. 279--311, 1966.

\end{thebibliography}

\end{document}